\documentclass{aa}  

\usepackage{graphicx}
\usepackage{color}
\usepackage{txfonts}
\usepackage[]{hyperref}
\newcommand\sfrd{$\Sigma_{\rm SFR}$}
\newcommand\T{\rule{0pt}{2.6ex}}       
\newcommand\B{\rule[-1.2ex]{0pt}{0pt}} 

\begin{document}

\title{Calibrating the relation of low-frequency radio continuum to star formation rate at 1 kpc scale with LOFAR}

\titlerunning{Low-frequency radio to SFR relation at 1 kpc scale}

\author{V. Heesen,\inst{1} E. Buie~II,\inst{2} CJ~Huff,\inst{2} L.~A.~Perez,\inst{2} J.~G.~Woolsey,\inst{2} D.~A.~Rafferty,\inst{1} A.~Basu,\inst{3} R.~Beck,\inst{4} E.~Brinks,\inst{5} \\ 
C.~Horellou,\inst{6} E.~Scannapieco,\inst{2} M.~Br\"uggen,\inst{1} R.-J.~Dettmar,\inst{7} K.~Sendlinger,\inst{7} B.~Nikiel-Wroczy\'{n}ski,$^{8}$ \\
K.\,T. Chy\.zy,\inst{8} P.~N.~Best,\inst{9} G.H.~Heald,\inst{10} \and R.~Paladino\inst{11}}
         
\authorrunning{V. Heesen et al.}

\institute{University of Hamburg, Hamburger Sternwarte, Gojenbergsweg 112, D-21029 Hamburg, Germany\\
\email{volker.heesen@hs.uni-hamburg.de}
\and School of Earth and Space Exploration, Arizona State University, P.O. Box 871404, Tempe, AZ, 85287-1404, USA
\and Fakult\"at f\"ur Physik, Universit\"at Bielefeld, Postfach 100131, D-33501 Bielefeld, Germany
\and Max-Planck-Institute f\"ur Radioastronomie, Auf dem H\"ugel 69, D-53121 Bonn, Germany
\and School of Physics, Astronomy and Mathematics, University of Hertfordshire, Hatfield AL10 9AB, UK
\and Chalmers University of Technology, Dept. of Space, Earth and Environment, Onsala Space Observatory, SE-439 92 Onsala, Sweden
\and Astronomisches Institut der Ruhr-Universit\"at Bochum, D-44780 Bochum, Germany
\and Astronomical Observatory, Jagiellonian University, ul. Orla 171, 30-244 Krak\'ow, Poland
\and SUPA, Institute for Astronomy, Royal Observatory, Blackford Hill, Edinburgh EH9 3HJ, UK
\and CSIRO Astronomy and Space Science, PO Box 1130, Bentley WA 6102, Australia
\and INAF/Istituto di Radioastronomia, via Gobetti 101, I-40129 Bologna, Italy}

   \date{Received month DD, YYYY; accepted month DD, YYYY}

 
  \abstract
   {Radio continuum (RC) emission in galaxies allows us to measure star formation rates (SFRs) unaffected by extinction due to dust, of which the low-frequency part is uncontaminated from thermal (free--free) emission.}
   {We calibrate the conversion from the spatially resolved 140 MHz RC emission to the SFR surface density ($\Sigma_{\rm SFR}$) at 1 kpc scale. Radio spectral indices give us, by means of spectral ageing, a handle on the transport of cosmic rays using the electrons as a proxy for GeV nuclei.}
   {We used recent observations of three galaxies (NGC 3184, 4736, and 5055) from the LOFAR Two-metre Sky Survey (LoTSS), and archival LOw-Frequency ARray (LOFAR) data of NGC~5194. Maps were created with the facet calibration technique and converted to radio $\Sigma_{\rm SFR}$ maps using the Condon relation. We compared these maps with hybrid $\Sigma_{\rm SFR}$ maps from a combination of \emph{GALEX} far-ultraviolet and \emph{Spitzer} 24 $\mu\rm m$ data using plots tracing the relation at the highest angular resolution allowed by our data at $1.2\times 1.2$-kpc$^2$ resolution.}
{The RC emission is smoothed with respect to the hybrid $\Sigma_{\rm SFR}$ owing to the transport of cosmic-ray electrons (CREs) away from star formation sites. This results in a sublinear relation $(\Sigma_{\rm SFR})_{\rm RC} \propto [(\Sigma_{\rm SFR})_{\rm hyb}]^{a}$, where $a=0.59\pm 0.13$ (140~MHz) and $a=0.75\pm 0.10$ (1365~MHz). Both relations have a scatter of $\sigma = 0.3~\rm dex$. If we restrict ourselves to areas of young CREs ($\alpha > -0.65$; $I_\nu \propto \nu^\alpha$), the relation becomes almost linear at both frequencies with $a\approx 0.9$ and a reduced scatter of $\sigma = 0.2~\rm dex$. We then simulate the effect of CRE transport by convolving the hybrid $\Sigma_{\rm SFR}$ maps with a Gaussian kernel until the RC--SFR relation is linearised; CRE transport lengths are $l=1$--5~kpc. Solving the CRE diffusion equation, assuming dominance of the synchrotron and inverse-Compton losses, we find diffusion coefficients of $D=(0.13$--$1.5)~\times 10^{28}~\rm cm^2\,s^{-1}$ at 1~GeV.}
{A RC--SFR relation at $1.4$~GHz can be exploited to measure SFRs at redshift $z \approx 10$ using $140$ MHz observations.}


   \maketitle
%
\begin{table*}
\centering
\caption{Properties of the sample galaxies. \label{tab:sample}}
\begin{tabular}{l cccc cc ccccc}
\hline
Galaxy & $i$ & $d_{25}$ & $D$ & $M_B$ & Type & Nucleus & $\rm SFR$ & $r_{\star}$ & $\log_{10}(\Sigma_{\rm SFR})$ & $\log_{10}(M_{\rm tot})$ & $\varv_{\rm rot}$ \\
& ($\degr$) & (arcmin) & (Mpc) & (mag) &  & & $(\rm M_{\sun}\, yr^{-1})$ & (kpc) & $({\rm M_{\sun}\, yr^{-1}\, kpc^{-2}})$ & ($\rm M_{\sun}$ ) &  ($\rm km\, s^{-1}$) \\
(1) & (2) & (3) & (4) & (5) & (6) & (7) & (8) & (9) & (10) & (11) & (12)\\
\hline
NGC~3184 & 16 & $7.41$  & $11.1$ & $-19.92$ & SBc & H\,{\sc ii} & $0.90$ & $10.7$ & $-2.62$ & $11.09$ & 210 \\
NGC~4736 & 41 & $7.76$  & $4.7$  & $-19.80$ & Sab & LINER       & $0.48$ & $4.1$  & $-2.07$ & $10.48$ & 156 \\
NGC~5055 & 59 & $11.75$ & $10.1$ & $-21.12$ & Sbc & T2          & $2.12$ & $16.5$ & $-2.63$ & $11.17$ & 192 \\
NGC~5194 & 20 & $7.76$  & $8.0$  & $-21.04$ & SBc & H\,{\sc ii} & $3.13$ & $15.1$ & $-2.39$ & $11.00$ & 219 \\
\hline
\end{tabular}
\flushleft{{\bf Notes.} Data are from \citet{walter_08a} and \citet{leroy_08a} unless otherwise noted.
\emph{Columns} (1) galaxy name; (2) inclination angle; (3) optical diameter measured at the $25~\rm mag\,arcsec^{-2}$ isophote in $B$-band; (4) distance; (5) absolute $B$-band magnitude; (6) galaxy morphological classification; 
(7) optical classification of the nuclear spectrum, from \citet{ho_97a} where $\rm Sy = Seyfert$ and $\rm T = transition$ object between H\,{\sc ii} nuclei and LINERs; (8) SFR based on the hybrid FUV + 24~$\mu$m conversion that has an uncertainty of $0.3$~dex; (9) radius of the actively star forming disc (within the last $\approx$100~Myr), estimated from the radial extent of the RC emission; (10) SFR surface density is $\Sigma_{\rm SFR}={\rm SFR}/(\pi r_{\star}^2)$ with an uncertainty of $0.3$~dex; (11) total mass $M_{\rm tot}=0.233\times 10^{10}r_{25} \varv_{100}^2~{\rm M_{\sun}}$, where $r_{\rm 25}$ is the radius of the galaxy estimated from $d_{25}$ using the distance $D$, and $\varv_{100}$ is the rotation speed in units of $100~\rm km\, s^{-1}$, with an uncertainty of $0.1$~dex; (12) maximum rotation speeds in the flat part of the rotation curve with typical uncertainties of $\pm 3~\rm km\, s^{-1}$.}
\end{table*}

\section{Introduction}
Radio continuum (RC) emission in galaxies emerges from two distinct processes: 
thermal free--free (bremsstrahlung) and non-thermal (synchrotron) radiation.
Both are related to the formation of massive stars. Ultraviolet (UV) radiation from massive stars ionises the interstellar medium (ISM), which gives rise to thermal bremsstrahlung emission. Studies of the origin of non-thermal RC emission have found that supernova
shocks from core collapse supernovae accelerate protons, heavier nuclei,
and electrons; non-thermal RC emission dominates at frequencies below 15~GHz; at higher
frequencies, such as a few 10~GHz, free--free emission dominates with a possible
contribution from spinning dust \citep[][]{scaife_10a}. The highly energetic electrons, known as cosmic-ray electrons (CREs), spiral
around interstellar magnetic field lines, thereby emitting highly linearly polarised synchrotron emission.
The relation between the  RC luminosity of a galaxy and its star formation rate
(SFR), henceforth referred to as the RC--SFR relation, is due to the interplay of star formation, CREs, and magnetic fields.

The global, integrated RC--SFR relation is very
tight, as \citet{heesen_14a} have shown: using the relation of
\citet{condon_92a} and converting $1.4$-GHz radio luminosities into radio derived
SFRs, these authors found agreement within 50 per cent with state-of-the-art star
formation tracers, such as far-ultraviolet (FUV), H~$\alpha$, and mid- or far-infrared (MIR; FIR) emission.
An even tighter agreement can be achieved if the radio spectrum is integrated over a wide frequency range \citep[`bolometric radio luminosity';][]{tabatabaei_17a}.
Moreover, these authors found that the radio luminosity is a non-linear function of the SFR,
as predicted by the model of \citet[][see Section~\ref{subsec:non_calorimetric_rc_sfr_relation} for details]{niklas_97a}.

These findings highlight that once properly calibrated, RC
can be used as an unobscured star-formation tracer in dusty, high-redshift galaxies \citep{beswick_15a} if the relation also holds at low frequencies. This is the case if magnetic fields at high redshifts are sufficiently strong to ensure that synchrotron losses of CREs dominate over inverse-Compton (IC) losses against the cosmic microwave background \citep{schleicher_13a}. Studies of the low-frequency RC--SFR relation have now become possible with the advent of the LOw-Frequency ARray \citep[LOFAR;][]{vanHaarlem_13a}. \citet{gurkan_18a} found that the integrated RC--SFR relation in the H-ATLAS field requires a broken power law to be described accurately. If fitted with a single power law, the relation between the 150 MHz RC luminosity and the SFR is $L_{150}\propto {\rm SFR}^{1.07}$, which is slightly super-linear. \citet{calistro_rivera_17a} studied the redshift evolution of the 150-MHz RC luminosity as function of FIR luminosity (RC--FIR relation). They showed the redshift evolution to be a potentially important factor when calibrating the usefulness of radio as a star formation tracer. \citet{chyzy_18a} found that the 150-MHz RC--FIR relation also holds in galaxies of the local Universe and has a similar scatter as for the $1.4$ GHz relation.

Low-frequency RC spectra of galaxies can be modified by additional mechanisms in comparison to GHz frequencies. Free--free absorption by thermal electrons can cause \ion{H}{II} regions to become optically thick at low frequencies, leading to a spectral turnover \citep{heesen_18b}. Relativistic bremsstrahlung and especially ionisation losses can also be much more important, particularly in starburst galaxies \citep{murphy_09a}. Furthermore, synchrotron self-absorption and Razin effects can further suppress the radio emission from dense ISM regions and produce spectral turnovers below 10\,MHz \citep{lacki_13a}. Starburst galaxies, such as M~82, become optically thick at 150~MHz when using spatially resolved observations \citep{chyzy_18a}. Since nuclear starbursts are similar, albeit on a smaller scale, a spatially resolved study allows us to separate these effects and the contribution from active galactic nuclei (AGNs) from our results. 


Calibrating the RC--SFR relation requires a thorough understanding of the
physical foundation that gives rise to the relation in the first
place. Measuring the non-linearity of the synchrotron--SFR relation and its potential dependence on galaxy type is crucial. Low-frequency observations are particularly useful because they allow
us to study the dominating synchrotron emission in galaxies, which is largely free from
the contribution of thermal emission. Furthermore, low-frequency
observations are ideally suited for large-area surveys, providing us with
statistically meaningful samples. This work exploits the recently
improved imaging capabilities of LOFAR to build up such a sample and study the
physics behind the RC--SFR relation. With spatially resolved observations we can explore whether we see a flattening of the RC--SFR relation in areas of concentrated star formation. Such a flattening would be hinting at a depression of RC intensities such as expected for free--free absorption.

Free--free absorption is one of the largest caveats in using low-frequency RC observations as a star formation tracer, particularly in starburst galaxies. \citet{adebahr_13a} found a spectral turnover for the integrated spectrum of M\,82 at 300~MHz, and even higher, at 600~MHz, for the starburst nucleus. Similarly, \citet{kapinska_17a} found the 500 pc nuclear starburst region in NGC~253 best described by an internally free--free absorbed synchrotron spectrum with a turnover frequency of 230~MHz. Clearly, these two galaxies represent the more extreme cases in the local Universe and their starburst nuclei have SFR surface densities well in excess of $1~\rm M_{\sun}\,yr^{-1}\,kpc^{-2}$, hence are factor of 10 or higher than what is usually referred to as starburst galaxies.

Low-frequency RC observations are particularly appealing to study star formation history across cosmic times. At redshift $z\approx 10$, the rest-frame $1.4$ GHz RC emission can be detected as 140 MHz emission. Since most deep large RC surveys with the next generation of radio telescopes will be performed at frequencies of 1--2~GHz, the $1.4$ GHz RC--SFR relation will be further established. This is for instance the case for planned surveys with the Square Kilometre Array (SKA) and its precursors, such as Australian Square Kilometre Array Pathfinder (ASKAP) Evolutionary Map of the Universe \citep[EMU;][]{norris_11a} and MeerKAT International GHz Tiered Extragalactic Exploration \citep[MIGHTEE;][]{jarvis_16a}. Similar surveys are planned with the next generation Very Large Array (ngVLA), the upgraded Giant Metrewave Telescope (uGMRT), and Multi-Element Radio Linked Interferometer Network (e-MERLIN).


This paper is an exploratory study of the low-frequency RC--SFR relation in four nearby galaxies, three of which (NGC~3184, 4736, and 5055) were observed as part of the LOFAR Two-metre Sky Survey \citep[LoTSS;][]{shimwell_17a}. The LoTSS is a deep 120--168~MHz imaging survey that will eventually cover the entire northern sky and reach an rms noise of $100~\mu\rm Jy\,beam^{-1}$ at an angular resolution of 5~arcsec. In addition, we added one galaxy that was observed previously \citep[NGC~5194 (M~51);][]{mulcahy_14a}, for which we reduced the data using the latest strategy. The galaxies were chosen from the SIRTF Nearby Galaxies Survey sample \citep[SINGS;][]{kennicutt_03a}, for which $1365$-MHz maps from the Westerbork Synthesis Radio Telescope exist \citep[WSRT--SINGS survey;][]{braun_07a}. As in \citet{heesen_14a}, we use the combined \emph{GALEX} 156-nm far-UV and \emph{Spitzer} $24$-$\mu$m MIR maps from \citet{leroy_08a} as our reference SFR surface density maps, in the following designated as hybrid $\Sigma_{\rm SFR}$ maps. The reasoning behind this choice is that the FUV data trace O and B stars, so they can be used as a star formation tracer as long as the obscuration by dust can be corrected for. See Table~\ref{tab:sample} for a summary of the properties of our sample galaxies.

This paper is organised as follows. In Section~\ref{sec:observations}, we present our observation strategy and data reduction techniques, including a comparison of our new map of NGC~5194 with the previously published map (Section~\ref{subsec:dde_vs_ddi}). Section~\ref{sec:rc_sfr_relation} describes our results for the RC--SFR relation; subsections are devoted to the morphology (Section~\ref{subsec:morphology}), dependency on the radio spectral index (Section~\ref{subsec:radio_spectral_index}), and spatially resolved relation (Section~\ref{subsec:regions_by_regions_study}). In Section~\ref{sec:cosmic_ray_transport}, we study the transport of cosmic rays, conducting a smoothing experiment in Section~\ref{subsec:cosmic_ray_diffusion_length} and  applying a diffusion model in Section~\ref{subsec:cosmic_ray_diffusion_model}. We discuss our results in Section~\ref{sec:discussion}, before we conclude in Section~\ref{sec:conclusions}. In the main part of the paper, we present the maps of NGC~5194 with the remaining galaxies presented in Appendix~\ref{app:maps}.


\section{Observations and data reduction}
\label{sec:observations}

Our new High Band Antenna (HBA) observations with LOFAR were taken with the LoTSS observing strategy (frequency and calibrator) set-up, observing the LoTSS pointings, which were closest to our targets. Our targets are within $2\degr$ from the pointing centres, thus having a primary beam attenuation of less than 25 per cent. In brief, we used the HBA-dual inner mode to conduct 8 hr observations; the 48 MHz bandwidth (120--168~MHz) was split equally over two target pointings, bookended by 10 min flux-calibrator scans (i.e.\ $8.3$~h scans); 50 per cent of the time of each scan was spent on our targets. We stored the data at 16 channels per sub-band ($12.2~\rm kHz$ frequency resolution) and at 1~s time resolution. The archival observation of NGC~5194 was carried out slightly differently. The bandwidth of 48~MHz was equally distributed between 116 and 176~MHz, with one pointing on the target and one pointing on the flux calibrator. See Table~\ref{tab:observations} for a journal of the observations.

\begin{table}
\centering
\caption{Journal of the observations.\label{tab:observations}}
\begin{tabular}{lc}
\hline
\multicolumn{2}{c}{NGC~3184}\\
Observation ID & L369724\\
Observation date & 2015 Aug 15\\
Project & LC4\verb|_|037\\
LoTSS pointing & P153+42 \\
Distance to pointing centre & $1\fdg 1$ \\
Stations & 60 (46 CS and 14 RS)\\
Primary calibrator & 3C~196 (L369720)\\
\multicolumn{2}{c}{NGC~4736 (observation 1)}\\
Observation ID & L343254\\
Observation date & 2015 May 13/14 \\
Project & LC3\verb|_|008 \\
LoTSS pointing & P191+42 \\
Distance to pointing centre & $1\fdg 3$ \\
Stations & 62 (48 CS and 14 RS)\\
Primary calibrator & 3C~196 (L343250)\\
\multicolumn{2}{c}{NGC~4736 (observation 2)}\\
Observation ID & L350666\\
Observation date & 2015 Jul 16\\
Project & LC4\verb|_|034\\
LoTSS pointing & P191+40\\
Distance to pointing centre & $1\fdg 8$ \\
Stations & 58 (44 CS and 14 RS)\\
Primary calibrator & 3C~196 (L350662)\\
\multicolumn{2}{c}{NGC~5055}\\
Observation ID & L280982 \\
Observation date & 2015 Mar 19/20 \\
Project & LC3\verb|_|008\\
LoTSS pointing & P198+42 \\
Distance to pointing centre & $0\fdg 3$ \\
Stations & 62 (48 CS and 14 RS)\\
Primary calibrator & 3C~196 (L280978)\\
\multicolumn{2}{c}{NGC~5194 (M~51)}\\
Observation ID & L127444 \\
Observation date & 2013 Apr 22/23 \\
Project & LC0\verb|_|043\\
LoTSS pointing & N/A \\
Distance to pointing centre & $0\degr$ \\
Stations & 61 (48 CS and 13 RS)\\
Primary calibrator & 3C~295 (L127444)\\
\hline
\end{tabular}
\flushleft {\bf Notes.} CS = core station; RS = remote station.\\
\end{table}

\begin{figure*}
\centering
\includegraphics[width=\hsize]{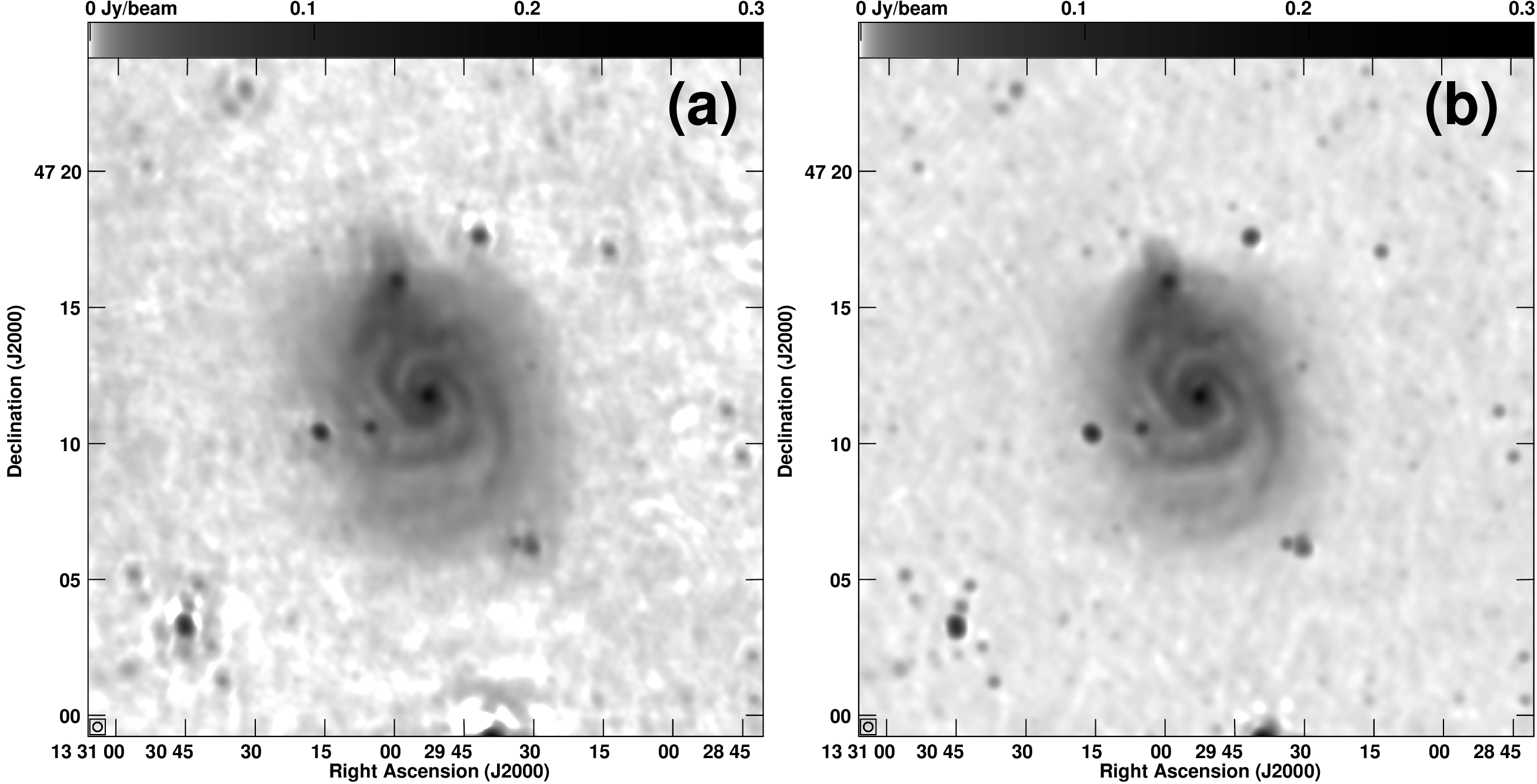} 
\caption{NGC~5194. (a) RC emission at 151~MHz, as derived from LOFAR HBA observations \citep{mulcahy_14a}. The map has been calibrated with direction-independent phase calibration only. (b) RC emission at 145~MHz, obtained from the same data set as (a) but calibrated with direction-dependent phase and amplitude calibration. Both maps are presented with a logarithmic transfer function and have an angular resolution of $20\times 20~\rm arcsec^2$, as indicated by the circle in the bottom left corner.}
\label{fig:n5194_dd_di}
\end{figure*}

\begin{table*}
\centering
\caption{Radio properties of sample galaxies. \label{tab:radio}}
\begin{tabular}{l c cc cc cc c cc}
\hline
Galaxy & FWHM & $\nu_1$ & $\nu_2$ & $\sigma_1$ & $\sigma_2$ & $S_1$ & $S_2$ & $\alpha$ & Int.\ Area & PA \\
& (arcsec) & \multicolumn{2}{c}{(MHz)} & \multicolumn{2}{c}{($\mu\rm Jy\, beam^{-1}$)} & \multicolumn{2}{c}{(Jy)} &  & (arcmin$^2$) & ($\degr$) \\
(1) & (2) & (3) & (4) & (5) & (6) & (7) & (8) & (9) & (10) & (11)\\
\hline
NGC~3184 & $18.6$ & $142$ & $1365$ & $130$ & $35$ & $0.389\pm 0.012$ & $0.087\pm 0.002$ & $-0.66\pm 0.02$ & $3.3\times 3.3$ & 179\\
NGC~4736 & $19.1$ & $140$ & $1365$ & $200$ & $42$ & $0.800\pm 0.024$ & $0.301\pm 0.006$ & $-0.43\pm 0.02$ & $3.0\times 2.5$ & 296\\
NGC~5055 & $18.6$ & $144$ & $1365$ & $110$ & $36$ & $2.082\pm 0.063$ & $0.416\pm 0.008$ & $-0.72\pm 0.02$ & $5.6\times 3.7$ & 102\\
NGC~5194 & $17.1$ & $145$ & $1365$ & $130$ & $32$ & $6.922\pm 0.208$ & $1.402\pm 0.028$ & $-0.71\pm 0.02$ & $6.5\times 5.8$ & 172\\
\hline
\end{tabular}
\flushleft{{\bf Notes.} \emph{Columns} (1) galaxy name; (2) angular resolution, referred to as the full width at half maximum (FWHM) of the circular synthesized beam; (3+4) observing frequencies $\nu_1$ and $\nu_2$; (5+6) rms map noises $\sigma_1$ and $\sigma_2$, at $\nu_1$ and $\nu_2$, respectively; (7+8) integrated flux densities, $S_1$ and $S_2$, at $\nu_1$ and $\nu_2$, respectively; (9) integrated radio spectral index between $\nu_1$ and $\nu_2$; (10) major and minor axes dimensions of the elliptical integration area; (11) position angle of the galaxy's major axis from \citet{walter_08a}.}
\end{table*}

The data were reduced with the facet calibration technique, which mitigates the direction-dependent effects of the ionosphere and beam response that impact low-frequency RC observations with aperture arrays, such that images close to the thermal noise level could be obtained \citep{van_weeren_16a,williams_16a}. First, the $(u,v)$ data are calibrated with direction-independent methods using the {\small PREFACTOR} pipeline \citep{degasperin_18a}.\footnote{\href{https://github.com/lofar-astron/prefactor}{https://github.com/lofar-astron/prefactor}} This pipeline first calibrates 3C~48 using the \citet{scaife_12a} flux densities, assuming a 
point-like source. From the resulting gain solutions the instrumental components are extracted, namely the station gain amplitudes and the phase variations due to the drift of the clocks of the LOFAR 
stations. The latter are separated from the variations due to the changing total electron 
content (TEC) of the ionosphere with the clock--TEC separation. Once determined, the instrumental calibration solutions are applied to the target data, which are then 
averaged to 10 s time resolution and  two channels per sub-band frequency resolution (channel 
width of $97.656~\rm kHz$). The data are calibrated in phase only using the Global Sky Model~\citep[GSM;][]{scheers_11a}, which is a compilation of sources from the VLA Low-frequency Sky Survey Redux \citep[VLSSr;][]{lane_14a}, the Westerbork Northern Sky Survey \citep[WENSS;][]{rengelink_97a}, and the NRAO VLA Sky Survey \citep[NVSS;][]{condon_98a}.  With the direction-independent calibration applied, the $(u,v)$ 
data are inverted and deconvolved with a wide-field {\small CLEAN} algorithm. As a final step of {\small PREFACTOR}, the {\small CLEAN} components of all the sources within the 
$8\degr$ field of view (FOV) are subtracted from the $(u,v)$ data.

The residual, direction-independent calibrated $(u,v)$ data with all sources subtracted, together with the subtracted model and solutions of the phase-only calibration, are then used as the input for the direction-dependent facet calibration, for which we used the {\small FACTOR} pipeline (Rafferty et al.\ 2018, in preparation).\footnote{\href{https://github.com/lofar-astron/factor}{https://github.com/lofar-astron/factor}} The FOV was divided into approximately 20 facets around calibrator regions with integrated 167 MHz flux densities (of the full facet) in excess of $0.3$~Jy.  Of those, facets in excess of $0.8$~Jy were processed one at a time, beginning with the brightest facet. The facet calibration technique allows us to track and correct for the direction-dependent effects of the ionosphere of the Earth (effectively the `seeing' at long radio wavelengths) and the station beam response by first self-calibrating on the calibrator region of a facet and then using the solutions to update the model for the full facet, which in turn is used to update the residual $(u,v)$ data. 
In the first step of the calibration, fast, 10 s phase solutions are determined in small chunks of $\approx$2~MHz bandwidth to correct for the positional change and distortion of sources. In the second step, slow, tens-of-minutes amplitude solutions are used to track the variation of the apparent flux density of a source. The target facets were corrected using the solution of a nearby facet.

The direction-dependent calibrated $(u,v)$ data were imported into the Common Astronomy Software Applications \citep[{\small CASA};][]{mcmullin_07a} and inverted and deconvolved with the MS--MFS {\small CLEAN} algorithm \citep{rau_11a}. We fitted for the frequency dependence of the skymodel ($\rm nterms=2$) and used angular scales of up to the size of the galaxy processed. We used Briggs weighting, setting the robust parameter between $0.2$ and $0.5$ and the value was adjusted to match the angular resolution of the WSRT maps. This resulted in maps with an effective central frequency ranging from 140 to 145~MHz with angular resolutions between 17 and 19 arcsec.\footnote{Angular resolutions in this paper are referred to as the full width at half maximum (FWHM).} In the following we refer to these data as the 140 MHz LOFAR data; see Table~\ref{tab:radio} for the map properties. We found rms noise levels between 110 and 200~$\mu\rm Jy\,beam^{-1}$, which is in approximate agreement with the expected sensitivity for our observations.

We integrated flux densities in our maps within ellipses encompassing the 3$\sigma$ contour lines. We checked the flux densities and found them in good agreement with the 7C catalogue (within 10 per cent), except for NGC~3184 for which there is no entry. Hence, we did not apply any correction for the well-known station calibration beam error \citep{hardcastle_16a}.


\subsection{Comparison with direction-independent calibration}
\label{subsec:dde_vs_ddi}
For NGC~5194, we have a map that was processed with direction-independent calibration by \citet{mulcahy_14a}. It is instructive to see the improvement that comes from the direction-dependent calibration technique using the facet calibration technique. In Fig.~\ref{fig:n5194_dd_di}, we present the comparison of the two maps, which have been convolved to $20\times 20~\rm arcsec^2$ resolution. The direction-independent map has an rms noise of 300--400~$\mu\rm Jy\,beam^{-1}$ in the area surrounding the galaxy, whereas the new map has a noise of 180--200~$\mu\rm Jy\,beam^{-1}$, hence the improvement is almost a factor of two. The advantage of the new map is also that the distribution of the noise is much more uniform across the map. Furthermore, sidelobes surrounding the brighter unresolved sources are significantly improved. We checked by fitting Gaussians to these sources such that the new map has a slightly improved resolution by about 1--2~arcsec, compared to the old map. Hence, this comparison shows that improvement using the direction-dependent calibration is significant even when the ionosphere is fairly quiescent as has been the case for this observation.

\section{RC--SFR relation}
\label{sec:rc_sfr_relation}
\subsection{Morphology}
\label{subsec:morphology}

Rather than working in observed flux density units, we make the assumption that the RC emission is entirely due to recent star formation. This was explored first by \citet{condon_92a}, who, based on some simple assumptions, predicted the following relation \citep[see][for details]{heesen_14a}:
\begin{equation}
  \frac{\rm SFR_{RC}}{\rm M_\sun\, yr^{-1}} =
0.75\times10^{-21} \left ( \frac{L_{1.4\,\rm GHz}}{\rm W\,Hz^{-1}}\right ).\end{equation}
Condon's relation assumes a Salpeter initial mass function (IMF) to extrapolate from the massive stars ($M > 5~\rm M_{\sun}$) that show up in the RC to that of all stars formed ($0.1 < M/{\rm M}_{\sun} < 100$). The hybrid \sfrd\  maps of \citet{leroy_08a}, which we employ as our reference, are based on a broken power-law IMF as described in \citet{calzetti_07a}, so their derived SFRs are a factor of $1.59$ lower than using the Salpeter IMF. We have thus scaled Condon’s relation in this paper accordingly.

Condon's relation is close to the result of the derivation of \citet{murphy_11a}, who used the empirical RC--FIR relation to derive SFRs that are 15 per cent lower. Condon’s relation can be generalised in two ways. First it can be scaled to any other frequency when one assumes a constant radio spectral index. We use  a radio spectral index of $\alpha = -0.8$ ($I_{\nu}\propto \nu^{\alpha}$), which is the total (i.e.\ including non-thermal and thermal emission) radio spectral index of galaxies at GHz frequencies \citep{tabatabaei_17a}. This may change below 1~GHz, where the radio spectral index of galaxies may flatten to approximately $-0.6$ \citep{chyzy_18a}, but in this paper we assume a simple power law without this further complication. Second, we can convert Condon's relation to a spatially resolved relation, thereby relating the SFR surface density with the RC intensity rather than luminosity. Using equation~(3) in \citet{heesen_14a}, we find
\begin{eqnarray}
\frac{(\Sigma_{\rm SFR})_{\rm RC}}{\rm M_{\sun}\,yr^{-1}\,kpc^{-2}} & = & 3.31 \times 10^3 \left(\frac{\nu}{1.4~\rm GHz}\right )^{0.8} \nonumber  \\
& &\times \left (\frac{\rm FWHM}{\rm arcsec}\right )^{-2} \frac{I_\nu}{\rm Jy\,beam^{-1}}.
\end{eqnarray}
This quantity is defined in the plane of the sky so that we do not have to deal with projection effects. Alternatively, we would have to correct the radio maps to face-on and do the same correction for the SFR maps. The correction factor is fairly small ($0.5<\cos(i)<1.0$) for our sample and does not make much difference in the analysis anyway, which is performed in log--log plots. There is also the subtle effect that the beam is elongated in the plane of the galaxy along the minor axis. This is not corrected for when we use rectangular regions, such as we do, as they sample a larger dimension along the minor axis. This would have to be corrected for by choosing regions with a smaller dimension along the minor axis. We chose to not do this. First, because the angular resolution of our maps is not sufficient to allow for it; second, \citet{heesen_14a} showed that results are not very sensitive to spatial resolution, comparing in that work $0.7$ and $1.2$~kpc spatial resolutions.

\begin{figure*}
\centering
\includegraphics[width=\hsize]{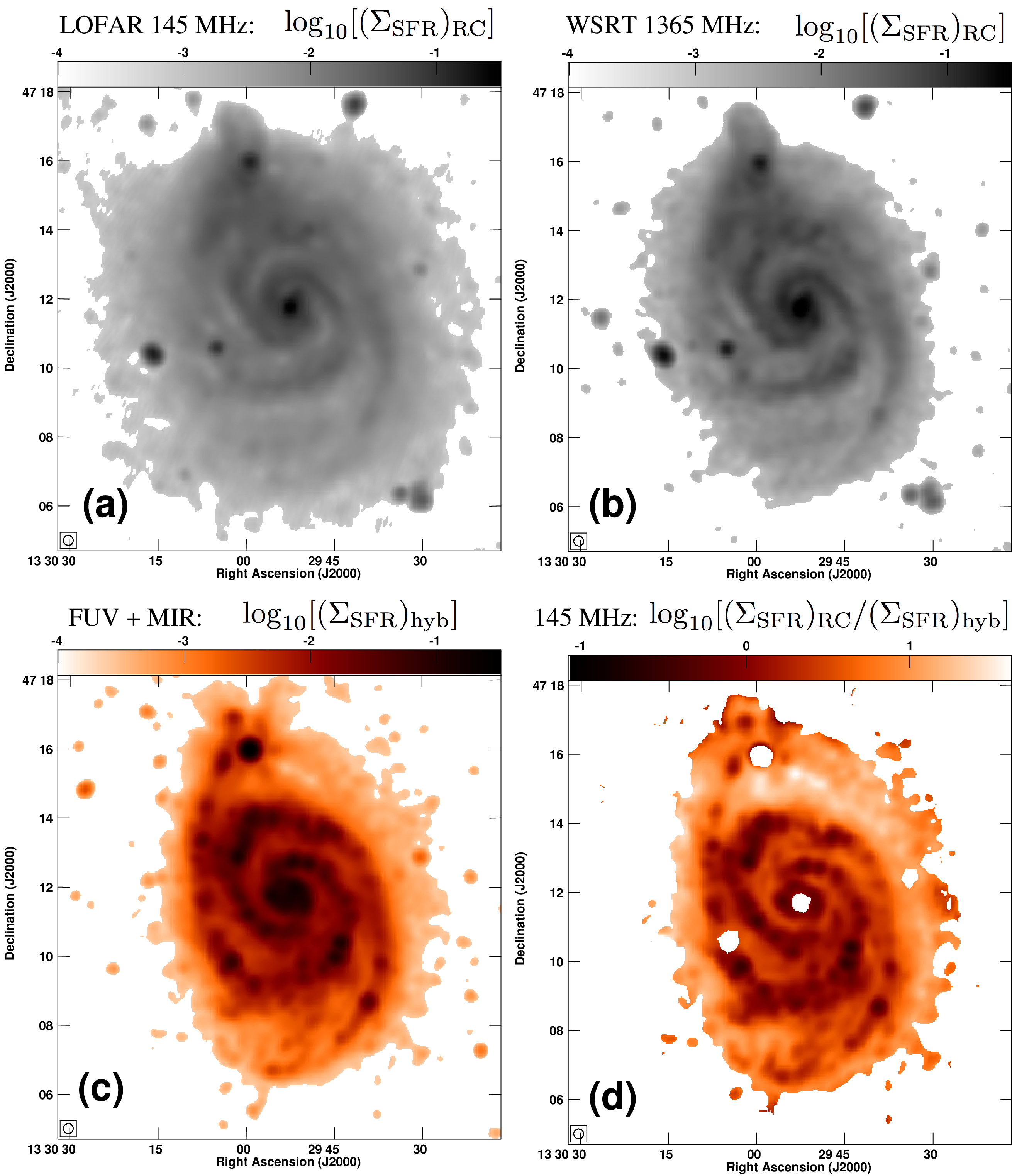} 
\caption{NGC~5194. (a) RC emission at 145~MHz, as derived from  LOFAR HBA observations. The intensities were converted into a map of the radio SFR surface density, $(\Sigma_{\rm SFR})_{\rm RC}$, using the $1.4$ GHz relation of Condon scaled with a radio spectral index of $-0.8$. This map is shown at a logarithmic stretch ranging from $10^{-4}$ to $3\times 10^{-1}~\rm M_{\sun}\,yr^{-1}\,kpc^{-2}$. (b) Same as (a), but using a $1365$-MHz map from WSRT--SINGS. (c) Hybrid SFR surface density map, $(\Sigma_{\rm SFR})_{\rm hyb}$, derived from a linear superposition of \emph{GALEX} 156 nm FUV and \emph{Spitzer} 24 $\mu$m MIR emission, presented as inverted heat colour scale. (d) Ratio, $\Re$, of the LOFAR $(\Sigma_{\rm SFR})_{\rm RC}$ map divided by the hybrid $(\Sigma_{\rm SFR})_{\rm hyb}$ map. The map is shown at logarithmic stretch using the heat colour scale, ranging from $10^{-1.1}$ to $10^{1.6}$. Areas that are light are radio bright, whereas dark areas are radio dim when compared with the hybrid $\Sigma_{\rm SFR}$-map. All maps have been convolved to a circular Gaussian beam with a resolution of $17.1\times 17.1$~arcsec$^2$. The representation of the beam is shown in the bottom left corner of each panel. Panels (a)--(c) show unmasked maps, whereas panel (d) shows the area after masking background sources and the AGN-contaminated central area. In all panels, a 3$\sigma$ cut-off has been applied.}
\label{fig:n5194_lofar}
\end{figure*}

The resulting radio \sfrd\  maps both of LOFAR and WSRT for NGC~5194 are presented in Figs~\ref{fig:n5194_lofar}(a) and (b), respectively. For comparison, we show in Fig.~\ref{fig:n5194_lofar}(c) the hybrid \sfrd-map. Clearly, the LOFAR map extends further than both the WSRT and hybrid \sfrd\  maps, particularly along the minor axis (major axis position angle is $\rm PA=172\degr$). The galaxy is inclined at $i=20\degr$, such that we may see a radio halo in projection. This is at least suggested by the morphology, as the radio emission does not extend very prominently along the major axis. 

We notice a number of unresolved sources. Most of these sources are easily identified as background radio galaxies since they have no counterpart in the hybrid \sfrd\  map. There are a number of compact sources in the hybrid \sfrd\ map as well, but these are located in the spiral arms; hence, we assume that they are massive star formation regions. We masked unrelated background sources and applied the mask to all \sfrd\  maps. Furthermore, we excluded compact nuclear sources since they may be dominated by AGNs. We stress that these masked sources contribute only a small amount of flux to the radio maps (at most 15 per cent), therefore they are not influencing our results in a significant way. In Fig.~\ref{fig:n5194_lofar}(d), we show the ratio of the LOFAR radio to hybrid \sfrd\  map with the mask applied. The map looks very similar to the hybrid \sfrd\  map with the spiral arms clearly visible. The heat colour scale is inverted, such that the minimum ratio is found in the spiral arms and the maximum ratio is found in the outskirts of the galaxy; the inter-arm regions show intermediate values for the ratio. 


The LOFAR \sfrd\  map shows less variation between the spiral arms, the inter-arm regions, and the outskirts of the galaxy than the hybrid \sfrd\  map. The radio map is in a way a smoothed version of the hybrid map. This is usually ascribed to the diffusion of the CREs away from star formation regions over their lifetime \citep{murphy_08a,heesen_14a}. The diffusion length is expected to be frequency dependent because at lower frequencies CREs have lower energies and longer lifetimes, assuming that synchrotron and IC radiation losses dominate as they do outside of the dense, gaseous spiral arms \citep{basu_15a}. Figures~\ref{fig:n5194_lofar}(a) and (b) hint that this is indeed the case with the contrast of the LOFAR map being even lower than that of the WSRT map.

Before we conclude this section, we briefly discuss the findings for the other three sample galaxies, the maps of which can be found in Appendix~\ref{app:maps}. NGC~3184 (Figs~\ref{fig:n3184_lofar} and \ref{fig:n3184_spix}) is only little inclined, and thus nearly in a face-on position. The spiral arms in the hybrid \sfrd\  map has faint counterparts in the radio maps. The other two galaxies are moderately inclined, NGC 4736 (Figs~\ref{fig:n4736_lofar} and \ref{fig:n4736_spix}) and 5055 (Figs~\ref{fig:n5055_lofar} and \ref{fig:n5055_spix}), with $i=41\degr$ and $i=59\degr$, respectively. As in NGC~5194, we found that the radio \sfrd\  maps extend further along the minor axis than the hybrid maps, more so than expected from thin inclined discs, suggesting the existence of radio haloes. All three galaxies show the same behaviour, namely that the radio emission is a smoothed version of the hybrid \sfrd\  map, the LOFAR maps even more so than the WSRT maps. 

Notable features are a large (10~arcmin) radio galaxy south-west of NGC~4736 (Figs~\ref{fig:n4736_lofar}(a) and (b)), where the northern lobe overlaps slightly with the emission of the galaxy. The hybrid \sfrd\  map (Fig.~\ref{fig:n4736_lofar}(c)) shows a filamentary extension to the west, of which we detect no counterpart in the radio. This emission is spatially coincident with a spiral arm visible in H\,{\sc i} emission \citep{walter_08a}, and thus may be a tidal tail caused by past interaction. This feature connects to a second outer ring visible in H\,{\sc i}, which may be caused by the Lindblad resonance \citep{schommer_76a}. In NGC~5055, the WSRT map (Fig.~\ref{fig:n5055_lofar}(b)) shows two extensions south-east and north-west of the main body. We find no counterpart in the LOFAR map (Fig.~\ref{fig:n5055_lofar}(a)) and neither is there one in either the hybrid \sfrd\  map (Fig.~\ref{fig:n5055_lofar}(c)) or in a map of H\,{\sc i} emission. This galaxy has an extended, warped H\,{\sc i} disc \citep{battaglia_06a} and also an extended M\,83-like FUV disc \citep{thilker_07a}. However, their morphology is different, which has maxima north-east and south-west of the bright, inner disc, whereas the WSRT extensions lie on the perpendicular axis. Hence, we conclude that this emission may be an artefact of the data reduction and we exclude this part of the galaxy from further analysis.

\begin{figure*}
\centering
\includegraphics[width=\hsize]{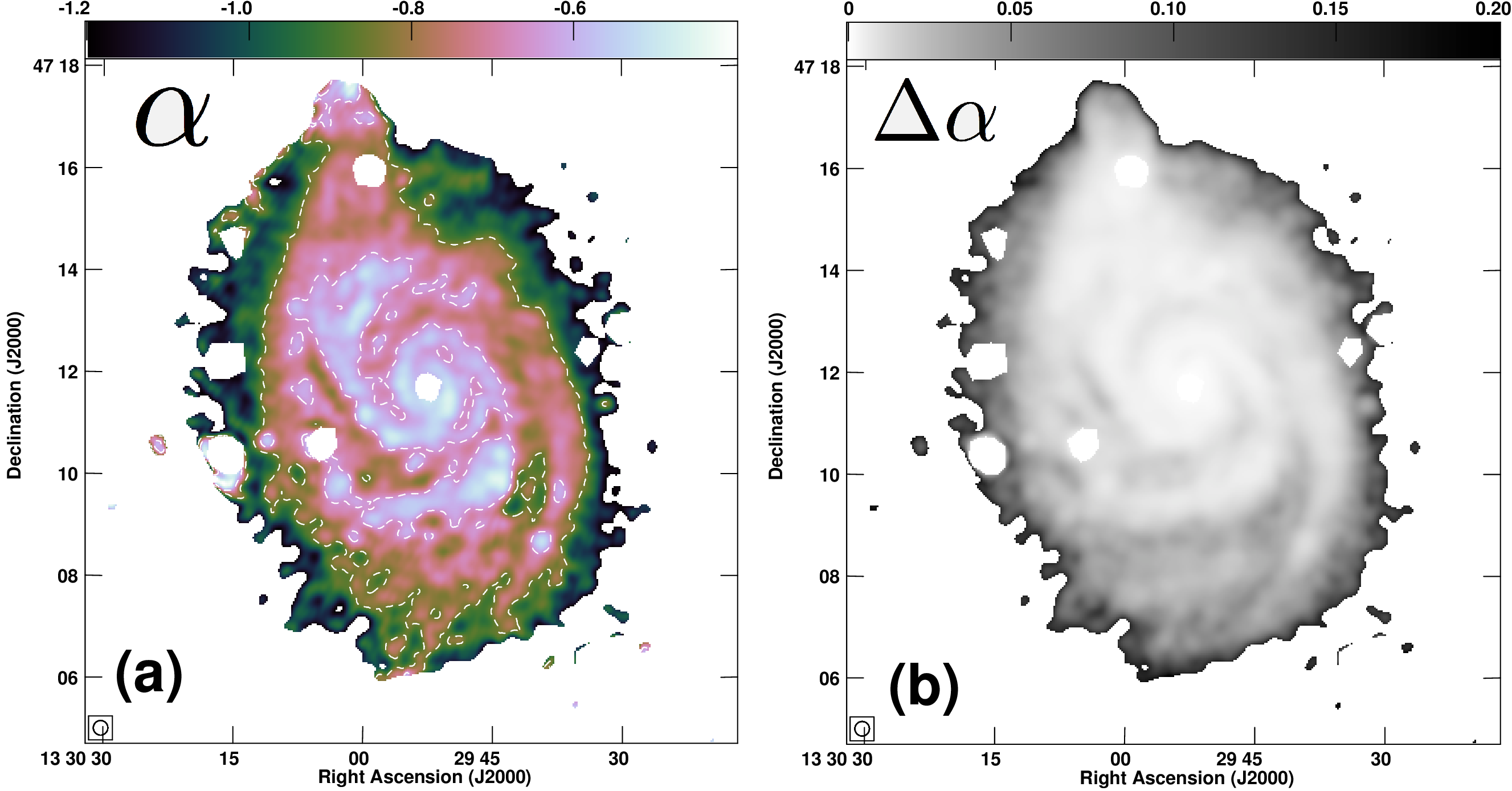} 
\caption{NGC~5194. (a) Radio spectral index distribution between 145 and 1365~MHz, presented in a cube-helix colour scale ranging from $-1.2$ to $-0.4$. Dashed contours are at $-0.85$ and $-0.65$, thus separating the galaxy in three zones. Areas with young CREs ($\alpha > -0.65$) are predominantly found in spiral arms; areas with CREs of intermediate age ($-0.85 \leq \alpha \leq -0.65$) are predominantly found in inter-arm regions; and areas with old CREs ($\alpha < -0.85$) are found in the galaxy outskirts. (b) Error of the radio spectral index distribution between 145 and 1365~MHz at logarithmic stretch, ranging from 0 to $0.2$. As can be seen, the spectral index error only becomes larger than $\pm 0.1$ in areas with $\alpha <
-0.85$. In both panels, the maps were convolved to a circular synthesised beam of $17.1\times 17.1~\rm arcsec^2$ resolution, which is outlined in the bottom left corner. A mask has been applied to background sources and the central regions of NGC~5194 and its companion galaxy, NGC~5195. A 3$\sigma$ cut-off was applied to both the 145 and 1365 MHz maps prior to combination.}
\label{fig:n5194_spix}
\end{figure*}

\begin{figure*}
\centering
\includegraphics[width=\hsize]{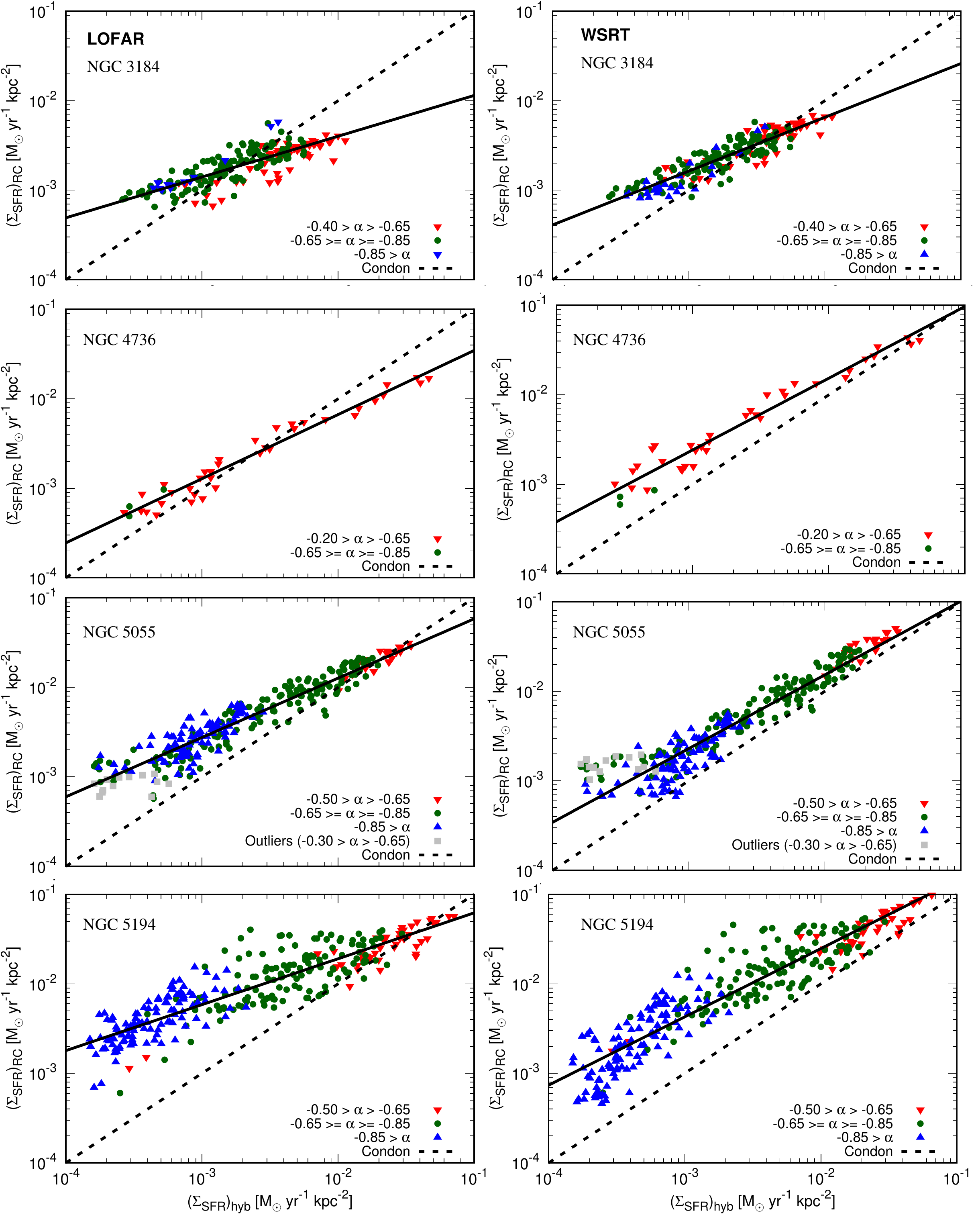} 
\caption{Plot of the individual galaxies, showing the spatially resolved RC--SFR ($(\Sigma_{\rm SFR})_{\rm RC}$--$(\Sigma_{\rm SFR})_{\rm hyb}$) relation. Each data point represents a $1.2\times 1.2~\rm kpc^2$ region that has been obtained from the hybrid $\Sigma_{\rm SFR}$ map (abscissa) and from the radio \sfrd\  map (ordinate). Shape and colour represent different radio spectral indices between 140 and 1365 MHz. Downward-pointing red triangles represent regions with young CREs ($-0.65 < \alpha < -0.20$); filled green circles represent regions with CREs of intermediate age ($-0.85 \leq \alpha \leq -0.65$); and upward-pointing blue triangles represent regions with old CREs ($\alpha < -0.85$). Solid black lines show the best-fitting relation and dashed lines show the Condon relation. The left panels show results for LOFAR 140 MHz and the right panels for WSRT 1365~MHz. A 3$\sigma$ cut-off was applied in all maps.}
\label{fig:pix_comb}
\end{figure*}
\begin{figure*}[!h]
\centering
\includegraphics[width=0.8\hsize]{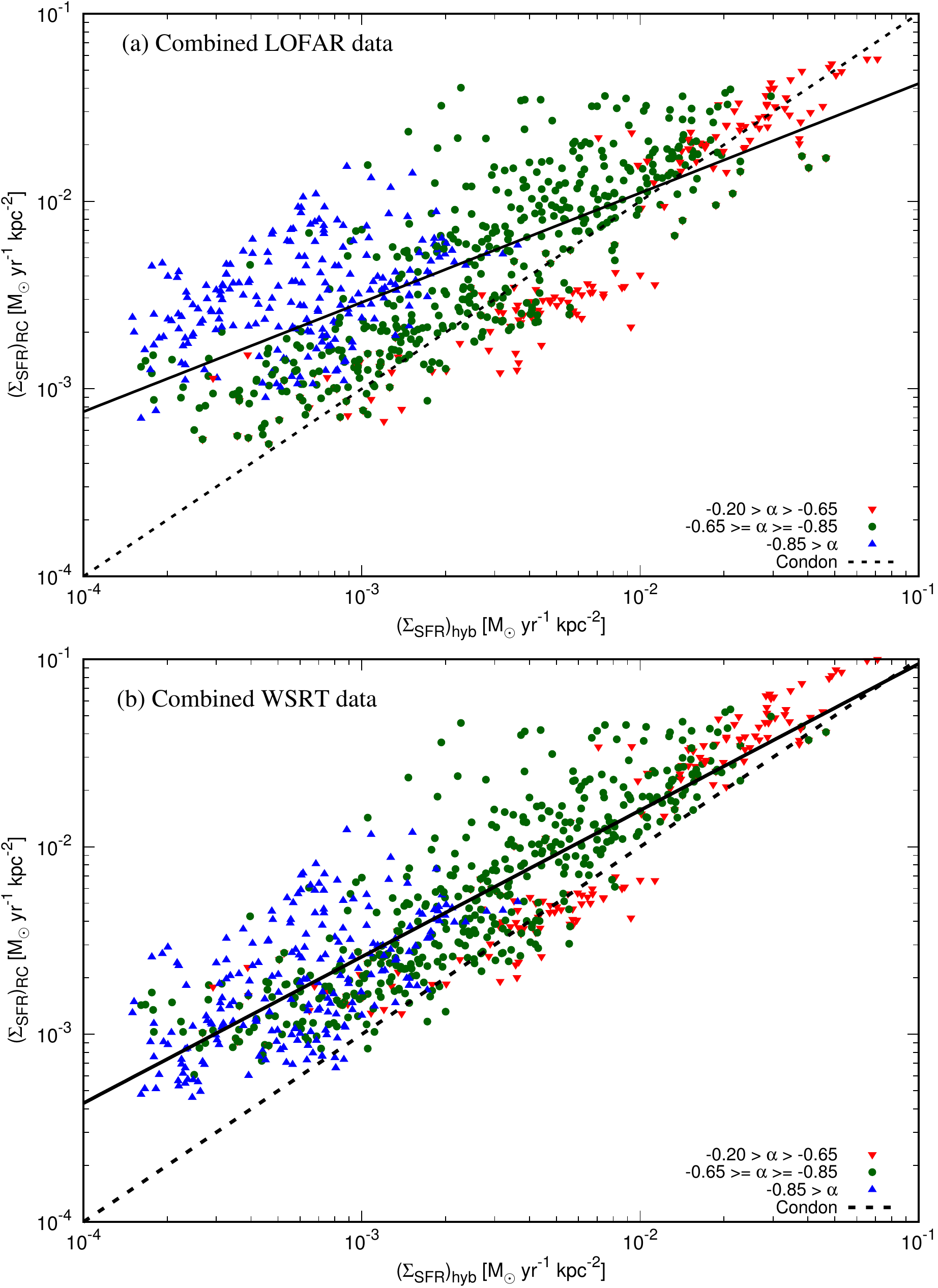} 
\caption{Plot of the combined data, showing the spatially resolved RC--SFR ($(\Sigma_{\rm SFR})_{\rm RC}$--$(\Sigma_{\rm SFR})_{\rm hyb}$) relation. Panel (a) shows results from LOFAR 140~MHz and panel (b) from WSRT 1365~MHz. Each data point represents a $1.2\times 1.2~\rm kpc^2$ region that has been obtained from the hybrid $\Sigma_{\rm SFR}$ map (abscissa) and from the radio $\Sigma_{\rm SFR}$ map (ordinate). Shape and colour represent different radio spectral indices between 140 and 1365 MHz. Downward-pointing red triangles represent regions with young CREs ($-0.65 < \alpha < -0.50$); filled green circles represent regions with CREs of intermediate age ($-0.85 \leq \alpha \leq -0.65$); and upward-pointing blue triangles represent regions with old CREs ($\alpha < -0.85$). Solid black lines show the best-fitting relation and dashed lines show the Condon relation. A 3$\sigma$ cut-off was applied in all maps.}
\label{fig:grand}
\end{figure*}

\subsection{Radio spectral index}
\label{subsec:radio_spectral_index}

In Fig.~\ref{fig:n5194_spix}, we present the 145--1365~MHz radio spectral index distribution in NGC~5194 with the corresponding error map; this is the total spectral index since we do not correct for thermal emission. The map is highly structured, where the galaxy can be roughly divided into three areas: (i) the spiral arms, where $\alpha>-0.65$; (ii) the inter-arm regions, where $-0.85 \leq \alpha \leq -0.65$; (iii) and the galaxies outskirts, where $\alpha < -0.85$. We varied the boundaries in the spectral index selection in order to best separate these regions based on the spectral index alone. The radio spectral index is nowhere flatter than $-0.5$, which is the expected injection index of young CREs from the theory of diffusive shock acceleration \citep{bell_78a}. This is supported observationally by the radio spectral index of supernova remnants, which is $-0.5\pm 0.2$ \citep{reynolds_12a}. However, without a fully sampled radio spectrum from the MHz to the GHz regime, we cannot be too sure whether the CREs are indeed young. Alternatively, they could be old, so their intrinsic spectral index may be $-0.8$, but free--free absorption suppresses the 140-MHz data point, turning it into an $\alpha$ of $-0.5$. Such a scenario is still possible and we can only exclude thermal self-absorption with additional observing frequencies. 

We find a similar situation in the other sample galaxies. The two galaxies with well-defined spiral arms, NGC~3184 and 5055, show  good agreement between the radio spectral index distribution and the location of the spiral arms, even though the contrast is not quite as pronounced as in NGC~5194. This is because these two galaxies show less prominent spirals arms in the hybrid \sfrd\  maps as well as in optical maps. In NGC~3184, the spectral index is everywhere steeper than $-0.4$ and in NGC~5055 steeper than $-0.5$, again in agreement with a CRE injection spectrum. NGC~3184 does not have any area, where the spectral index is steeper than $-0.85$. This could be in part caused by a sensitivity limitation, since this galaxy has the joint lowest hybrid \sfrd\  value and thus lowest RC surface brightness. NGC~5055 has emission with a steep spectral index, in particular along the minor axis. It also shows some flat spectral indices in the halo, but as discussed before, we believe that this is due to spurious emission in the WSRT map; emission that we discard for the following analysis. This is supported by the fact that our 140 MHz RC map shows a more extended halo than at 1365~MHz, but with a completely different morphology hinting at a radio halo that we see in projection as discussed in Section~\ref{subsec:morphology} \citep[see also][for a 333-MHz map of this galaxy]{basu_12a}.

With regards to the spectral index distribution, NGC~4736 is different from the other galaxies since the spectral index is fairly flat. The maximum local spectral index is $-0.2$, therefore it becomes questionable whether we see the CRE injection spectral index. The thermal fraction of the RC emission in the `starburst ring' is so high that it flattens the spectral index even at these low frequencies \citep{basu_12a}. Furthermore, possible explanations for the flat spectral indices are ionisation and bremsstrahlung losses, which both depend on neutral (atomic and molecular) gas density. Since this galaxy has high gas surface densities of 50--100~$\rm M_{\sun}\, pc^{-2}$ within a radius of 100~arcsec \citep{leroy_08a}, where we observe mainly the flat spectral indices, such an explanation seems likely. This is in fair agreement with the $>$$200~\rm M_{\sun}\,pc^{-2}$ that \citet{basu_15a} suggested for areas of $\alpha > -0.5$, but for spectral indices between 330~MHz and $1.4$~GHz. Clearly, this is a strong function of frequency, such that a threshold lower by a factor of two can be expected for our lower observing frequency.


\begin{table*}
\centering
\caption{Spatially resolved RC--SFR relation, the $(\Sigma_{\rm SFR})_{\rm RC}$-- $(\Sigma_{\rm SFR})_{\rm hyb}$ relation, at $1.2$-kpc spatial resolution. \label{tab:regions}}
\begin{tabular}{l ccc ccc c}
\hline
Galaxy & $a_{1}$ & $b_{1}$ & $\sigma_{1,1200}$ & $a_{2}$ & $b_{2}$ & $\sigma_{2,1200}$ & $\rm FWHM_{\rm 1200}$ \\
& & & (dex) & & & (dex) & (arcsec) \\
(1) & (2) & (3) & (4) & (5) & (6) & (7) & (8)\\
\hline
NGC~3184           & $0.46\pm 0.02$ & $-1.49\pm 0.06$ & $0.12$ & $0.60\pm 0.02$ & $-0.98\pm 0.05$ & $0.10$ & $22.29$ \\
NGC~4736           & $0.72\pm 0.02$ & $-0.74\pm 0.07$ & $0.10$ & $0.80\pm 0.03$ & $-0.21\pm 0.08$ & $0.12$ & $52.64$ \\ 
NGC~5055           & $0.68\pm 0.02$ & $-0.53\pm 0.04$ & $0.15$ & $0.84\pm 0.02$ & $-0.13\pm 0.05$ & $0.16$ & $24.50$ \\
NGC~5194           & $0.51\pm 0.02$ & $-0.69\pm 0.04$ & $0.20$ & $0.76\pm 0.02$ & $-0.08\pm 0.05$ & $0.22$ & $30.93$ \\
$\rm Combined^{a}$     & $0.59\pm 0.02$ & $-0.77\pm 0.05$ & $0.31$ & $0.79\pm 0.02$ & $-0.23\pm 0.04$ & $0.25$ & $-$     \\
$\rm Young~CRE^{b}$ & $0.93\pm 0.03$ & $-0.21\pm 0.06$ & $0.21$ & $0.91\pm 0.03$ & $-0.04\pm 0.06$ & $0.21$ & $-$     \\
\hline
\end{tabular}
\flushleft{{\bf Notes.} \emph{Columns} (1) data plotted; (2+3) best-fitting parameters for equation~(\ref{eq:rc-sfr}) for $\nu_1\approx 140~\rm MHz$; (4) standard deviations for $\nu_1$; (5+6) best-fitting parameters for equation~(\ref{eq:rc-sfr}) for $\nu_2=1365~\rm MHz$; (7) standard deviations for $\nu_2$; (8) angular resolution given as FWHM of the map, which is equivalent to a projected spatial resolution of $1.2~\rm kpc$; (a) relation for the plot of the combined data points from the galaxy sample as presented in Fig.~\ref{fig:grand}; (b) relation for young CREs using only data points for which $\alpha > -0.65$ as presented in Fig.~\ref{fig:grand_sl1}.}
\end{table*}

\begin{figure*}
\centering
\includegraphics[width=\hsize]{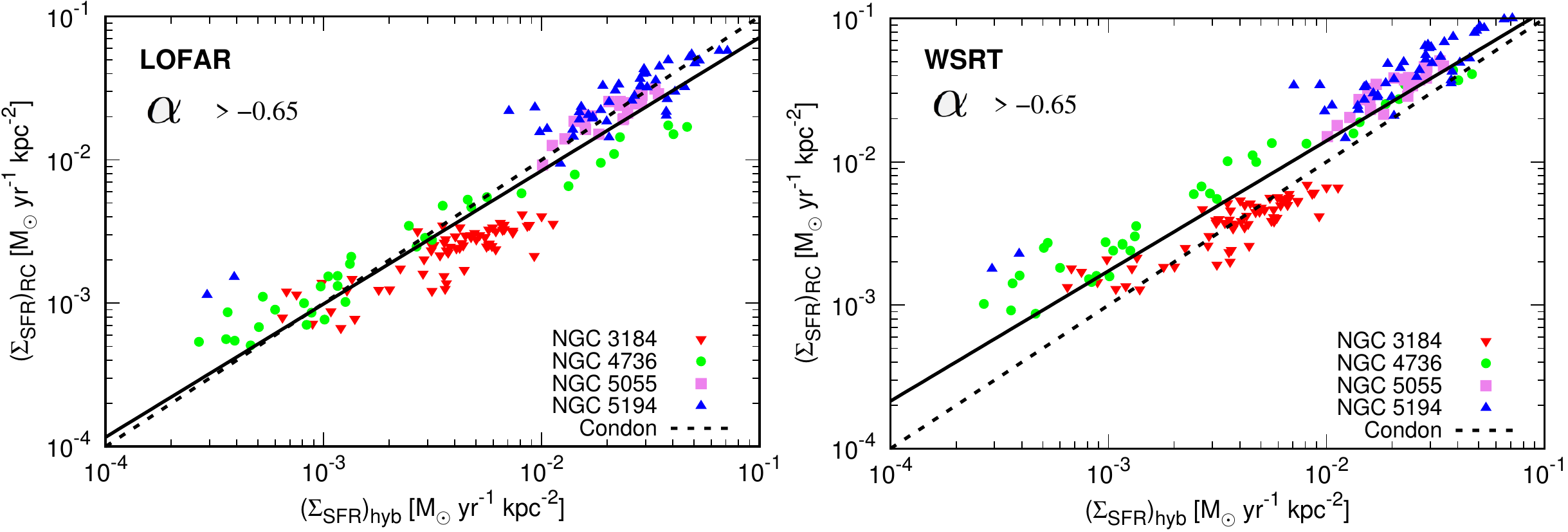} 
\caption{Plot of the combined data, showing the spatially resolved RC--SFR ($(\Sigma_{\rm SFR})_{\rm RC}$--$(\Sigma_{\rm SFR})_{\rm hyb}$) relation, for data points with spectral indices $\alpha>-0.65$ only. Panel (a) shows results from LOFAR 140~MHz and panel (b) from WSRT 1365~MHz. Each data point represents a $1.2\times 1.2~\rm kpc^2$ region that has been obtained from the hybrid $\Sigma_{\rm SFR}$ map (abscissa) and from the radio \sfrd\  map (ordinate). Shape and colour represent the different galaxies. Solid black lines show the best-fitting relation and dashed lines show the Condon relation. A $3\sigma$ cut-off was applied in all maps.}
\label{fig:grand_sl1}
\end{figure*}

\subsection{Spatially resolved RC--SFR relation}
\label{subsec:regions_by_regions_study}

In this section, we study the spatially resolved RC--SFR relation, the $(\Sigma_{\rm SFR})_{\rm RC}$--$(\Sigma_{\rm SFR})_{\rm hyb}$ relation, at the resolution limit of our data. We measure \sfrd\ values averaged in regions of $1.2\times 1.2~\rm kpc^2$ size in the radio \sfrd\  maps from LOFAR and WSRT and in the hybrid \sfrd\  maps. Prior to this, we convolved the maps with a Gaussian to a resolution (FWHM)
that corresponds to a projected linear scale of $1.2$~kpc. We applied 3$\sigma$ cut-offs in all maps before creating the regions-by-regions plots. For each region, the radio spectral index between 140 and 1365~MHz was computed. We present the resulting plots for our four sample galaxies in Fig.~\ref{fig:pix_comb}. In each plot the best-fitting least-squares linear relation (using the Marquardt--Levenberg algorithm) is presented as well as the prediction from Condon's relation. The least-squares fitting was done in log--log space, fitting the function
\begin{equation}
        \log_{10} [(\Sigma_{\rm SFR})_{\rm RC}] = a \log_{10}[(\Sigma_{\rm SFR})_{\rm hyb}] + b.
    \label{eq:rc-sfr}
\end{equation}
Hence, in this notation one can write\begin{equation}
(\Sigma_{\rm SFR})_{\rm RC} = 10^b [(\Sigma_{\rm SFR})_{\rm hyb}]^{a},
\end{equation}
where $a$ represents the power-law slope of the spatially resolved RC--SFR relation, when considering that the radio \sfrd\  map is directly proportional to the RC intensity, and $b$ is a constant offset. The resulting best-fitting parameters can be found in Table~\ref{tab:regions}. We find the slope of the relation is $0.46$--$0.72$ for LOFAR and $0.60$--$0.84$ for WSRT. This confirms our earlier result of sublinear slopes \citep{heesen_14a}. For each galaxy we find that the LOFAR slope is even flatter than for WSRT, such as we already hinted in the study of the morphology (Section~\ref{subsec:morphology}). 

A second result is that the offset from Condon's relation is a function of the radio spectral index. Areas with steep spectral indices (green and blue data points) are relatively speaking `radio bright', compared with what is expected from Condon's relation. Data points representing areas with young CREs (red data points) are in good agreement with the Condon relation. Interestingly this is not a function of the hybrid \sfrd. In NGC~4736, which has the highest spatially resolved RC--SFR slope, data points with young CREs can be found from $3\times 10^{-4}$ to $6\times 10^{-2}~\rm M_{\sun}\,yr^{-1}\,kpc^{-2}$. These data points are in good agreement with Condon's relation (within a factor of 2), in particular for LOFAR.\footnote{For WSRT the agreement can be improved, depending on the normalisation of the Condon relation.} On the other hand, we find in NGC~5055 and 5194 young CREs only in areas where the hybrid \sfrd\  map exceeds approximately $10^{-2}~\rm M_{\sun}\,yr^{-1}\,kpc^{-2}$. These are also the areas where the agreement with Condon's relation is best. In NGC~3184 the spectral index separation is not so clear with all points clustering around the Condon relation.

\begin{figure*}
\centering
\includegraphics[width=0.8\hsize]{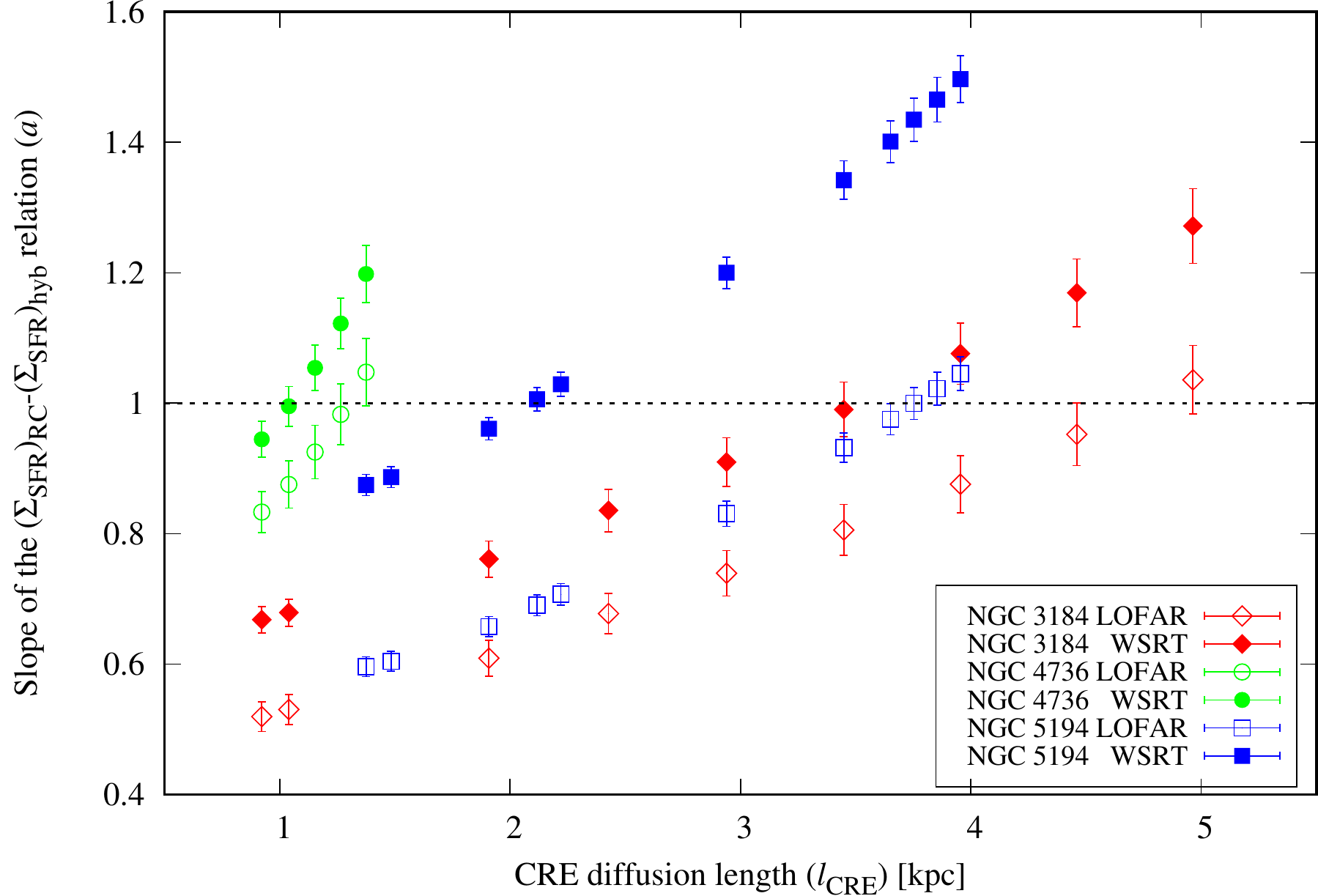} 
\caption{Smoothing experiment to measure the CRE diffusion length. Plotted is the slope of the $(\Sigma_{\rm SFR})_{\rm RC}$--$(\Sigma_{\rm SFR})_{\rm hyb}$ relation, $a$, as function of the CRE diffusion length $l_{\rm CRE}$. LOFAR 140 MHz data are shown as open and WSRT 1365 MHz data are shown as filled symbols. The three galaxies are represented by red (NGC~3184), green (NGC~4736), and blue (NGC 5194).}
\label{fig:convolution}
\end{figure*}

\begin{figure*}
\centering
\includegraphics[width=0.8\hsize]{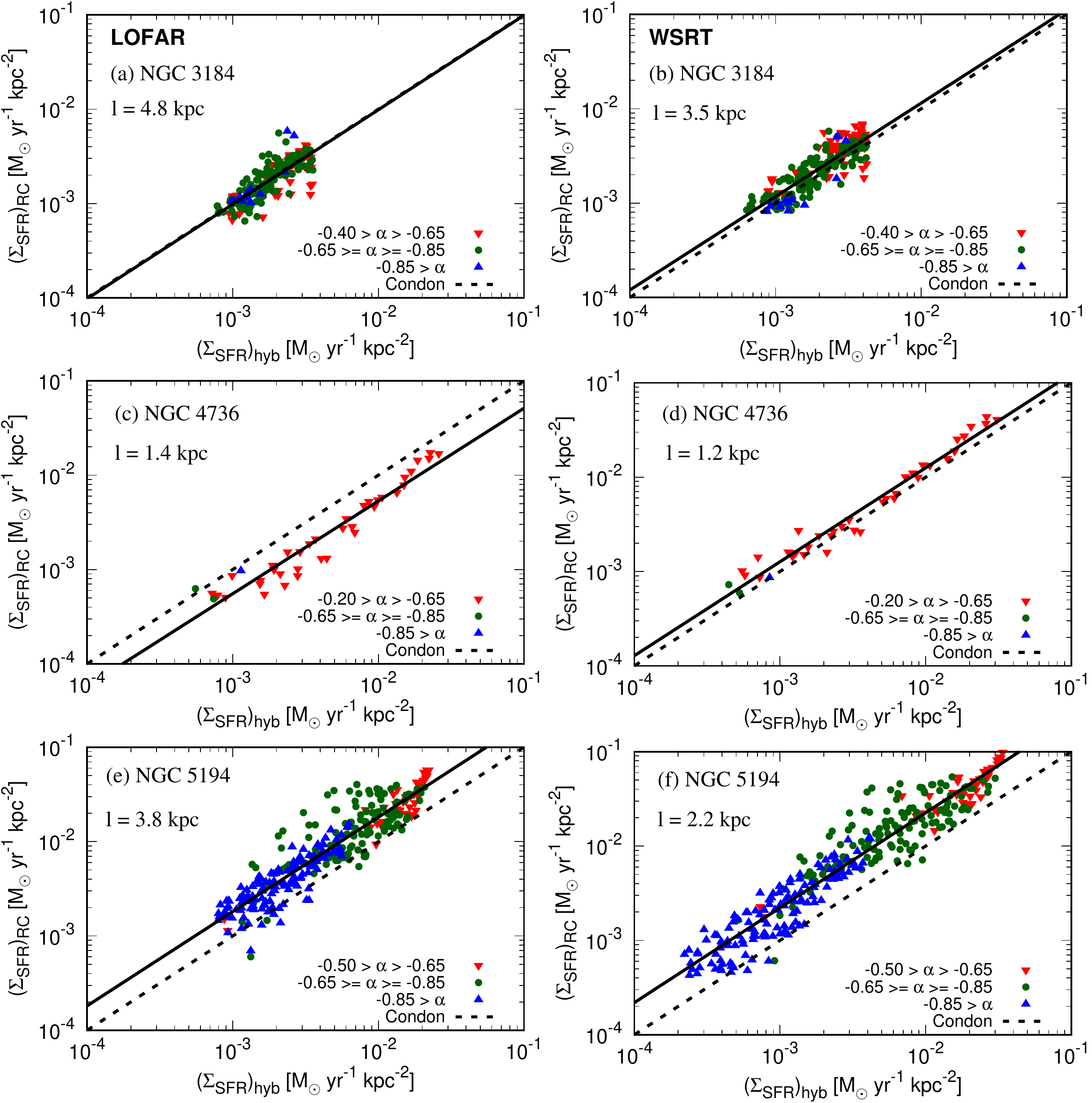} 
\caption{Linearised  $(\Sigma_{\rm SFR})_{\rm RC}$--$(\Sigma_{\rm SFR})_{\rm hyb}$ relation after convolving the $(\Sigma_{\rm SFR})_{\rm hyb}$ map with a Gaussian kernel to simulate the diffusion of CREs. Panels on the left-hand side show results for the 140 MHz LOFAR data and panels on the right-hand side results for the 1365 MHz WSRT data. The length of the Gaussian kernel, $l$ (=FWHM/2), is noted as well. Panels (a) and (b) show results for NGC~3184, panels (c) and (d) for NGC~4736, and panels (e) and (f) for NGC~5194. Each data point represents a $1.2\times 1.2~\rm kpc^2$ regions that has been obtained once from the hybrid $\Sigma_{\rm SFR}$ map (abscissa) and once from the radio $\Sigma_{\rm SFR}$ map (ordinate). Shape and colour represent different radio spectral indices between 145 and 1365 MHz. Downward-pointing red triangles represent regions with young CREs ($-0.65 < \alpha < -0.20$); filled green circles represent regions with CREs of intermediate age ($-0.85 \leq \alpha \leq -0.65$); and upward-pointing blue triangles represent regions with old CREs ($\alpha < -0.85$). Solid black lines show the best-fitting relation and dashed lines show the Condon relation. A $3\sigma$ cut-off was applied in all maps.}
\label{fig:linear}
\end{figure*}

In the next step, we combined the data from all four galaxies at a fixed linear scale of $1.2\times 1.2~\rm kpc^2$ in one plot for LOFAR and WSRT each, presented in Fig.~\ref{fig:grand}. As for the individual galaxies, we find that the spatially resolved RC--SFR relation has a slope of $0.59$ for LOFAR, which is smaller than for WSRT with $0.79$. We also find that the scatter of the LOFAR relation is with $0.3$~dex larger than for WSRT with $0.25$~dex. It now becomes even more strikingly apparent that the red data points are best in agreement with Condon's relation regardless of the hybrid \sfrd, whereas the green and blue data points lie predominantly above Condon's relation.

As the last step, we investigated this further by repeating a plot of the combined data, but only of data points with spectral indices $\alpha>-0.65$. This is shown in Figs~\ref{fig:grand_sl1} (a) and (b) for LOFAR and WSRT, respectively. Indeed, now all data points are in good agreement with Condon's relation. Most importantly, the relation is now much more linear with a slope of $0.93$ for LOFAR and $0.91$ for WSRT. This means that we can find a normalisation factor for a linear spatially resolved RC--SFR relation. Over three decades of hybrid \sfrd, we can find a relation that strays only $0.15$~dex from a linear relation. The best-fitting relations are also fairly tight with scatters of only $0.21$~dex both for LOFAR and WSRT. Hence, this expands significantly on earlier results in \citet{heesen_14a}, where it was suggested that only areas with hybrid \sfrd\  values in excess of approximately $10^{-2}~\rm M_{\sun}\,yr^{-1}\,kpc^{-2}$ can be recommended for the use of RC emission as a reliable star formation tracer. Our new LOFAR observations suggest that it is instead the radio spectral index that is the better discriminant.

Another approach was used by \citet{dumas_11a} and \citet{basu_12a} who manually selected arm and inter-arm regions, where the former is dominated by young CREs. According to \citet{basu_12a}, the arm regions of NGC~5194 reveal a slope of the spatially resolved RC--FIR relation of $1.0$, while the slope in the inter-arm regions is close to $2$, probably owing to strong synchrotron losses of CREs diffusing from the arms. \citet{basu_12a} showed that the arm regions in their 333 MHz maps of a sample of galaxies display a higher spatially resolved RC--FIR relation slope of $0.6\pm 0.1$ compared to the inter-arm regions that have a slope of $0.3\pm 0.1$. Two of their galaxies, NGC~4736 and 5055, are also part of our sample. They find RC--FIR slopes that are flatter than our results even when we consider only their arm regions whereas we fit the entire galaxies. Since their RC--FIR slopes at $1.4$ GHz are also lower than the RC--SFR slopes we measure, we conclude that the difference is due to the intrinsic difference between our hybrid \sfrd\  maps and the \emph{Spitzer} 70 $\mu$m FIR maps they used. Qualitatively at least, our results are in good agreement.

There is also some remaining scatter, for instance the data points of NGC~3184 are systematically below that of the other galaxies, meaning that this galaxy is radio weak. This can be caused by weak magnetic fields and escape of CREs from the galaxy. The latter can happen either through advection in a galactic wind or diffusive transport, the properties of which we investigate in the next section.


\section{Cosmic-ray transport}
\label{sec:cosmic_ray_transport}


\subsection{Smoothing experiment}
\label{subsec:cosmic_ray_diffusion_length}
In this section, we investigate the properties of the cosmic-ray transport. Because the CREs are injected into the ISM at star formation
sites and are transported away during their lifetime, the resulting CRE
distribution is a smoothed version of the \sfrd\  map \citep{bicay_90a, murphy_06a}. As we have seen in Section~\ref{subsec:regions_by_regions_study}, the 1 kpc scale RC--SFR relation is sublinear, where the slope of the relation is more shallow for LOFAR than for WSRT. The sublinear slopes have been reported before \citep{tabatabaei_13a,heesen_14a}, but the trend with frequency is new. Although this has been suggested before from the radio spectral index as separate parameter in the 1 kpc RC--SFR relation \citep{heesen_14a}, we can now measure the cosmic-ray transport length as a function of frequency and compare this with theory. 

First, we measure the transport length. This can be carried out by convolving the hybrid \sfrd\  map with a suitable kernel to linearise the spatially resolved RC--SFR relation \mbox{\citep{tabatabaei_13a,heesen_14a}}. The first choice that has to be made is the shape of the transport kernel. There are two different modes of transport of cosmic rays in galaxies: advection in a galactic wind and diffusion along magnetic field lines. In an earlier work of \citet{murphy_06a}, it was shown that a exponential kernel is a marginally better representation than a Gaussian kernel, even though their approach was slightly different; these authors studied the correlation with only either MIR or FIR emission using \emph{Spitzer} 24 and 70 $\mu\rm m$ emission. We expect advection to be only of importance in the halo and since we are studying emission mostly from the thin disc, we expect diffusion to be the dominant process. For such a case we expect that a Gaussian diffusion kernel is the correct approach \citep{heesen_16a}. 


In Fig.~\ref{fig:convolution}, we show the results of this smoothing experiment. We exclude NGC~5055 from the analysis, since the cosmic-ray transport process in this galaxy will be investigated in the future (Sendlinger et al. 2018, in preparation). The CRE diffusion was simulated by convolving the $(\Sigma_{\rm SFR})_{\rm hyb}$ map with a Gaussian kernel. In this way, the sublinear $(\Sigma_{\rm SFR})_{\rm RC}$--$(\Sigma_{\rm SFR})_{\rm hyb}$ relation can be linearised, corresponding to $a = 1$ as shown by the horizontal line in Fig.~\ref{fig:convolution}, providing us with a measurement of the CRE diffusion length. As we can see, we are indeed able to linearise the spatially resolved RC--SFR relation in all studied galaxies; the resulting linearised plots are presented in Fig.~\ref{fig:linear}. Now we define the length of the diffusion kernel, $l$, as the half width at half maximum ($l= \rm FWHM/2$) of the Gaussian convolution kernel applied to the hybrid \sfrd\  map. The CRE diffusion length can then be derived as
\begin{equation}
l_{\rm CRE}^2 = l^2 - l_{\rm beam}^2,
\label{eq:diffusion_length}
\end{equation}
where we correct for the smoothing owing to the limited resolution of our maps with $l_{\rm beam}=0.6$~kpc. Thus derived diffusion lengths can be found in Table~\ref{tab:cosmic_ray_diffusion}. For LOFAR, we find projected (in the plane of the sky) diffusion lengths between $1.3$ and $4.8$~kpc, and for WSRT between $1.0$ and $3.5$~kpc. In each galaxy, the diffusion length for LOFAR is larger than for WSRT with a minimum ratio of $1.2$ and a maximum ratio of $1.7$. We checked that the intrinsic resolution of the LOFAR and WSRT images are the same, which we expect, since we corrected the LOFAR image for the blurring effects of the ionosphere of the Earth with the direction-dependent calibration (Section~\ref{sec:observations}). We fitted unresolved sources with a 2D Gaussian function using {\small IMFIT} in {\small AIPS} and found agreement within 5--10 per cent (1--2~arcsec) of the fitted FWHM in the LOFAR and WSRT images. 
\begin{table*}
\centering
\caption{Properties of cosmic ray diffusion. \label{tab:cosmic_ray_diffusion}}
\begin{tabular}{l cc cc cc cc cc }
\hline
Galaxy & $B_0$ & $U_{\rm rad}/U_{\rm B}$ & $E_1$ & $E_2$ & $\tau_1$ & $\tau_2$ & $l_{\rm CRE1}$ & $l_{\rm CRE2}$ & $D_1$ & $D_2$ \\
& ($\mu\rm G$) &  & \multicolumn{2}{c}{(GeV)} & \multicolumn{2}{c}{(Myr)} & \multicolumn{2}{c}{(kpc)} & \multicolumn{2}{c}{($10^{28}~\rm cm^2\, s^{-1}$)}\\
(1) & (2) & (3) & (4) & (5) & (6) & (7) & (8) & (9) & (10) & (11) \\
\hline
NGC~3184 & $8.1$  & $0.16$ & $1.15$ & $3.57$ & $124\pm 31$ & $40\pm 10$ & $4.8\pm 0.3$ & $3.4\pm 0.2$ & $5.5\pm 1.6$ & $8.9\pm 2.5$\\ 
NGC~4736 & $10.3$ & $0.56$ & $1.01$ & $3.17$ & $74\pm 19$  & $24\pm 6$  & $1.3\pm 0.1$ & $1.0\pm 0.1$ & $0.6\pm 0.2$ & $1.4\pm 0.5$\\
NGC~5055 & $9.6$  & $0.22$ & $1.07$ & $3.29$ & $105\pm 26$ & $34\pm 9$  & N/A          & N/A          & N/A          & N/A\\
NGC~5194 & $12.2$ & $0.25$ & $0.95$ & $2.92$ & $70\pm 18$  & $23\pm 6$  & $3.8\pm 0.1$ & $2.1\pm 0.1$ & $5.9\pm 1.6$ & $5.8\pm 1.7$\\
\hline
\end{tabular}
\flushleft{{\bf Notes.} Entries with subscripts `$1$` and `$2$` are at observing frequencies $\nu_1\approx 140~{\rm MHz}$ and $\nu_2=1365~{\rm MHz}$, respectively. \emph{Columns} (1) galaxy name; (2) magnetic field strength in the disc estimated using {\small BFIELD} from \citet{beck_05a}. We assumed a path-length of 1~kpc, a proton-to-electron ratio of $K=100$, and non-thermal radio spectral index of $-0.8$, and energy equipartiton; (3) ratio of radiation energy density to magnetic field energy density. The first was estimated from the total infrared luminosities of \citet{galametz_13a}, including a contribution from the stellar population and the cosmic ray microwave background \citep[see][]{heesen_18a}; (4+5) CRE energies from equation~(\ref{eq:CRE_E}); (6+7) CRE lifetimes from equation~(\ref{eq:CRE_tau}); (8+9) CRE diffusion lengths (equation~(\ref{eq:diffusion_length})); analytic estimate of the CRE diffusion coefficients as defined in equation~(\ref{eq:diffusion_coefficient}).}
\end{table*}

We can calculate the diffusion coefficients using the following simplified equation first:
\begin{equation}
D = \frac{l_{\rm CRE}^2}{\tau},
\label{eq:diffusion_coefficient}
\end{equation}
where $\tau$ is the CRE lifetime due to synchrotron and IC radiation losses. This is the 1D case for anisotropic diffusion along magnetic field lines. For a 3D case for isotropic diffusion, the diffusion coefficients would be a factor of four lower. The CRE energy at the observing frequency $\nu$ can be calculated from
\begin{equation}
E({\rm GeV}) = \sqrt{\left (\frac{\nu}{16.1~{\rm MHz}}\right ) \left ( \frac{B_{\perp}}{\rm \mu G}\right )^{-1}},
\label{eq:CRE_E}
\end{equation}
where $B_{\perp}$ is the perpendicular magnetic field strength, which can be approximated as $B_{\perp}=\sqrt{2/3}\,B_0$ for an isotropic turbulent magnetic field, where $B_0$ is the total magnetic field strength in the disc plane. The CRE lifetime is
\begin{equation}
\tau = 8.352\times 10^9 \left (\frac{E}{\rm GeV} \right )^{-1} \left (\frac{B_{\rm \perp}}{\rm \mu G}\right )^{-2}\left (1+\frac{U_{\rm rad}}{U_{\rm B}}\right )^{-1}~{\rm yr}.
\label{eq:CRE_tau}
\end{equation}
Above, we have also included IC radiation losses, where $U_{\rm rad}$ is the radiation energy density including the interstellar radiation and cosmic microwave background. The ratio of synchrotron to IC losses is equivalent to the ratio of the magnetic energy density, $U_{\rm B}=B^2/(8\pi)$, to the radiation energy density (see Table~\ref{tab:cosmic_ray_diffusion} for resulting CRE energies, lifetimes, diffusion lengths, and diffusion coefficients). We find CRE energies between 1 and 4~GeV and lifetimes of approximately 100 and 30~Myr for LOFAR and WSRT, respectively. The resulting diffusion coefficients are between $0.6$ and $8.9~\times 10^{28}~\rm cm^2\,s^{-1}$. These values are in broad agreement with what has been found in external galaxies before \citep{berkhuijsen_13a,tabatabaei_13a} and also with the Milky Way value of $3\times 10^{28}~\rm cm^2\,s^{-1}$ \mbox{\citep{strong_07a}}. In one galaxy, NGC~5194, we find that the diffusion coefficient does not increase with energy, whereas it may in the other two galaxies. We return to discuss these findings in Section~\ref{dis:cosmic_ray_transport}.

\citet{murphy_08a} studied the transport of CREs in the same sample as we did. Interestingly, the resulting smoothing lengths have the same trend that we find with the smallest length in NGC~4736 and the largest length in NGC~3184. However, their lengths are a factor of two smaller than we find at 1365~MHz. We attribute this to the fact that they smoothed \emph{Spitzer} $70$~$\mu$m images, rather than hybrid \sfrd\  maps as we do. Since the FIR emission is sensitive to less warm dust than the $24~\mu\rm m$ MIR emission, the FIR emission is more spread out over the disc and less concentrated in spiral arms and other areas of recent star formation. This means that the $70$ $\mu \rm m$ map has to be less smoothed in order to resemble the radio map. This is in line with the findings of \citet{basu_12a} and our spatially resolved RC--SFR relation (Section~\ref{subsec:regions_by_regions_study}). Furthermore, \citet{murphy_08a} minimised the difference between the RC and FIR maps, whereas we linearised the RC--SFR relation  such that a difference can be expected.

\citet{mulcahy_14a} used scale-dependent wavelet transforms of the LOFAR 151~MHz and \emph{Spitzer} $70$~$\mu$m images of NGC~5194 and their cross-correlation spectra to measure the diffusion lengths of CREs. Their result of $1.45$~kpc is lower by a factor of $2.6$ compared to our result at 145~MHz (Table~\ref{tab:cosmic_ray_diffusion}). This discrepancy indicates that the cross-correlation coefficient $r_w$ of $0.75$ for measuring the diffusion length was too small. A choice of $r_w = 0.85$ would yield a diffusion length similar to our result.

\begin{figure}
\centering
\includegraphics[width=\hsize]{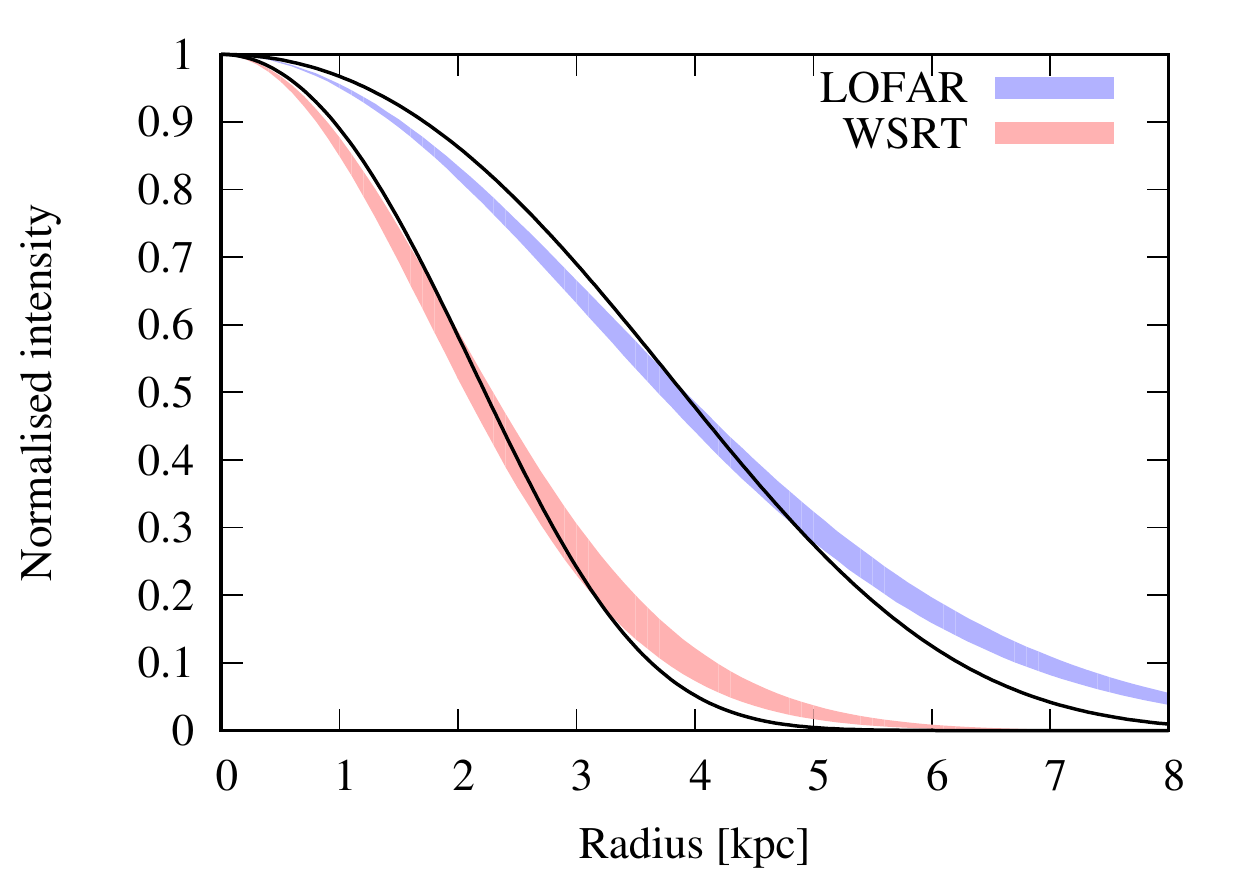} 
\caption{Diffusion kernels in NGC~5194. Normalised intensity as function of radius, i.e.\ distance to the star formation (CRE injection) region. The blue and red shaded areas show the measured diffusion kernels assuming a Gaussian distribution for LOFAR and WSRT, respectively. The areas indicate the range of uncertainties. Solid lines represent the best-fitting cylindrically symmetric cosmic-ray diffusion model.}
\label{fig:n5194_diff}
\end{figure}

\begin{table}
\centering
\caption{Best-fitting cosmic ray diffusion models. \label{tab:cosmic_ray_diffusion_models}}
\begin{tabular}{l c cc c}
\hline
Galaxy & $U_{\rm IRF}/U_{\rm B}$ & $D_0$ & $\mu$ \\
& & ($10^{28}~\rm cm^2\,s^{-1}$) &\\
(1) & (2) & (3) & (4)\\
\hline
NGC~3184 & $0.16$ & $0.9_{-0.3}^{+0.8}$    & $0.5$\T\B\\ 
NGC~4736 & $0.46$ & $0.13_{-0.04}^{+0.09}$  & $0.6$\B\\
NGC~5055 & $0.10$ & N/A & N/A\B \\
NGC~5194 & $0.18$ & $1.5_{-0.4}^{+1.2}$   & $0.0$\B\\
\hline
\end{tabular}
\flushleft{{\bf Notes.} NGC~5055 will be analysed separately in a forthcoming paper (Sendlinger et al. 2018, in preparation). \emph{Columns} (1) galaxy name; (2) ratio of the interstellar radiation field energy density (excluding the contribution from the cosmic microwave background) to the magnetic energy density; (3) 2D diffusion coefficient at 1~GeV, where the uncertainty interval is for 3D (lower) and 1D (upper bound); (4) energy dependence of the diffusion coefficient parametrised as $D=D_0 (E/{\rm GeV})^{\mu}$.}
\end{table}

\subsection{Cosmic-ray diffusion model}
\label{subsec:cosmic_ray_diffusion_model}

We now introduce a more physically motivated description of cosmic-ray diffusion. As before, we use the CREs as proxies for cosmic rays at GeV energies, from which the bulk of the cosmic-ray pressure originates. Our model relies on there being a steady state between the injection of CREs and energy losses;  we assume only synchrotron and IC radiation losses. Hence, for a steady-state we find
\begin{equation}
\Delta N(E) = \frac{1}{D} \left \lbrace\frac{\partial}{\partial E} [b(E) N(E)]\right \rbrace,
\end{equation}
where $b(E)$ are the CRE energy losses. This is equivalent to equation~(8) in \citet{heesen_16a}, except that we have now used the Laplace operator, such that we can extend this to a 2D and 3D case. This equation is solved numerically from the inner boundary where we assume that the cosmic-ray number density is $N(E,0) = N_0 E^{-\gamma}$, where $\gamma$ is the CRE energy injection index. The synchrotron emission is then found by convolving the CRE number density with the synchrotron spectrum of an individual CRE. We also convolved the radial intensity profile with the resolution of our radio maps $\rm FWHM=1.2~kpc$.

In the following we investigated three cases: (i) the 1D diffusion case, which would be appropriate for anisotropic diffusion along magnetic field lines, such as the spiral arm magnetic fields found in many late-type galaxies; (ii) the 2D diffusion case, such as expected if the CREs diffuse in the disc plane since differential rotation causes magnetic field lines to be predominantly aligned with the orientation of the disc; and (iii) the 3D diffusion case for an isotropic diffusion, such as expected for turbulent magnetic fields. Generally speaking, the 1D diffusion coefficients are the largest and the 3D coefficients are the smallest. Since realistic values are probably found in between the extremes, we used the 2D values and give the variation going from 1D to 3D as an error estimate. Intensity profiles are calculated for constant magnetic field strengths $B_0$ using the SPectral INdex Numerical Analysis of K(c)osmic-ray Electron Radio-emission ({\small SPINNAKER}).\footnote{\href{https://github.com/vheesen/Spinnaker}{https://github.com/vheesen/Spinnaker}} We parametrised the energy dependence of the diffusion coefficient as $D=D_0 (E/{\rm GeV})^{\mu}$. This is the expected behaviour since higher energy CREs have larger Larmour radii and, consequently, larger mean free paths, such that they scatter less often. From theory, depending on which kind of turbulence is assumed, $\mu$ is expected to be in the range from $0.3$ to $0.6$ \citep{strong_07a}, although such a relation may be seen only for CREs in excess of a few GeV \citep{recchia_16b}.


As an example, we show in Fig.~\ref{fig:n5194_diff} the resulting diffusion kernels for NGC~5194. Shaded areas show the Gaussian kernels that we found by linearising the spatially resolved RC--SFR relation in Section~\ref{subsec:cosmic_ray_diffusion_length} and solid lines show the best-fitting 2D diffusion model. It is obvious that the assumption of a Gaussian distribution is a fair representation of the modelled diffusion kernels, such that  we can now justify this earlier assumption. The resulting diffusion coefficients for the three modelled galaxies are presented in Table~\ref{tab:cosmic_ray_diffusion_models}. The diffusion coefficients are a factor of a few lower than the analytic estimate we made previously in Section~\ref{subsec:cosmic_ray_diffusion_length}. The reason is two-fold: first, the flat injection CRE energy index of $\gamma=2.2$, which is what we assume to be consistent with our earlier spectral index measurements such that the injection radio spectral index is $\alpha = -0.6$, means that the CRE spectrum is harder because relatively speaking there are more high-energy CREs. This partially compensates for the fact that CREs with the highest energies lose their energy fastest. Second, the use of a 2D symmetric diffusion reduces the coefficient further. For NGC~5194, the 1D best-fitting coefficient is only a factor of 2 lower than the analytic estimate. Using a CRE energy injection index of $\gamma=3.0$ removes the discrepancy almost entirely.

We find diffusion coefficients between $0.13$ and $1.5\times 10^{28}~\rm cm^2\,s^{-1}$ and energy dependencies between $\mu=0.0$ and $0.6$. This is in good agreement with theoretical predictions as well as from observations \citep{strong_07a}. Studies that used the RC smoothing technique found diffusion coefficients of the order $10^{28}~\rm cm^2\,s^{-1}$ \citep{murphy_08a,berkhuijsen_13a,tabatabaei_13a}. The energy dependence has so far been investigated only in NGC~5194 by \citet{mulcahy_16a}, using a model to describe the radial diffusion from inwards to outwards in the galaxy. We agree with their result that we do not see an energy dependence in this galaxy, but our diffusion coefficients are smaller by a factor of four. We discuss this aspect in more detail in Section~\ref{dis:cosmic_ray_transport}.


\section{Discussion}
\label{sec:discussion}

\subsection{Origin of the RC--SFR relation}
\label{subsec:non_calorimetric_rc_sfr_relation}

It has been realised early on \mbox{\citep{condon_92a}}
that for the non-thermal RC emission to be
related to the SFR,
(1) the CREs have either to emit all their energy within a
galaxy, such that the galaxy constitutes an electron calorimeter; or
(2a) the cosmic rays have to be in energy equipartition with the magnetic field and
(2b) there is a $B$--$\Sigma_{\rm SFR}$ or a $B$--$\rho$ relation \mbox{\citep{niklas_97a}}. The
relations of the second condition are closely interrelated via the
Kennicutt--Schmidt relation, where the SFR is a function of the gas mass or
density ($\rho$). 

Model (1) predicts a linear synchrotron--SFR relation, model (2) a non-linear relation and the index depends on whether or not energy equipartition is valid on the linear scale of investigation. Present-day observations of spiral galaxies seem to favour model (2), which is expected, since electron calorimetry might hold at best in starburst galaxies, but almost certainly not in low-mass dwarf irregular (dIrr) galaxies, which lose a large fraction of their CREs in galactic winds and
outflows \citep{chyzy_16a}. While massive starburst galaxies also have winds, their magnetic field strengths are sufficiently high, such that they are closer to a calorimeter, where synchrotron losses dominate over escape losses \citep{li_16a}. \citet{hindson_18a} showed that dwarf galaxies are radio weak in relation to their SFR, such that the relation can be
fitted by a single power law with $L\propto \rm SFR^{1.3}$, where $L$ is the RC luminosity.


Our finding of an almost linear spatially resolved RC--SFR relation at 1 kpc scale for young CREs (Section~\ref{subsec:regions_by_regions_study}), which is in good agreement with the relation by Condon, now corroborates the non-calorimetric origin of the RC--SFR relation. Since at low frequencies we observe only non-thermal RC emission, the RC--SFR relation has to hold for CREs with injection spectral indices as found in the spiral arms and other areas of star formation. Clearly, these areas are locally (1 kpc scale) not electron calorimeters as their spectral indices are in good agreement with the injection spectral index of young CREs. In contrast, areas that have steep spectral indices tend to lie above the Condon relation. These areas are closer to a calorimeter, although the situation is more complex because we may see only aged CREs since the local star formation is low and injection of young CREs can be almost neglected. What is certain is that calorimetry does not hold locally when observing young CREs, and some kind of relation between cosmic-ray and magnetic energy density has to exist. The most natural explanation is equipartition, where the strength of the magnetic field regulates the storage of the cosmic rays in the galaxy.

There are cases where the non-thermal RC--SFR relation breaks down, namely in dwarf galaxies \citep{hindson_18a}. We can now explain this as well. Since the cosmic-ray diffusion length can be a few kpc, which is comparable to the size of an entire dwarf galaxy, the CREs can easily leave the galaxy and thus lead to a violation of equipartition. Alternatively, the correlation breaks down if the SFR is too low to maintain the correlation between SFR and magnetic field strength \citep{schleicher_16a}.

Finally, we calculate what the predicted $B$--$\Sigma_{\rm SFR}$ relation has to be to explain our results. For areas of young CREs, we assume $\alpha = -0.6$, and the best-fitting average RC--SFR relation transposed to RC intensities, where $I_\nu \propto ((\Sigma_{\rm SFR})_{\rm hyb})^{0.92}$. Because the non-thermal RC luminosity scales with the magnetic field strength as $I_\nu \propto N_{\rm CR} B^{1-\alpha}$, where $N_{\rm CR}$ is the cosmic-ray number density and $\alpha$ the non-thermal radio spectral index, we find $B\propto ((\Sigma_{\rm SFR})_{\rm hyb})^{0.26}$. This is slightly flatter than what the integrated RC--SFR relations suggest, which tend to be steeper. Using a Kennicutt--Schmidt type law of star formation \citep{kennicutt_12a}, we can also calculate the $B$--$\rho$ relation. Using $(\Sigma_{\rm SFR})_{\rm hyb}\propto \rho^{n}$, with $n=1.4$ as appropriate for atomic gas, we find $B\propto \rho^{0.36}$. This is slightly lower than the relation predicted by turbulent magnetic field amplification (e.g. by dynamo action), which predicts $\beta\approx 0.5$ \citep{beck_96a,thompson_06a}, where the magnetic field strength is parametrised as $B\propto \rho^{\beta}$. A similar observed value of $\beta\approx 0.4$ was found in \citet{dumas_11a} for the spiral arm regions of NGC~5194 and $\beta\approx 0.7$ for the inter-arm regions, the former in fair agreement with our results.

The equipartition model of \citet{niklas_97a} predicts a slope $s$ of the RC--SFR relation of $s = (3-\alpha) \, \beta / n$. The observed slope of $\approx 0.9$ of the relation for young CREs ($\alpha \approx -0.6$) and assuming $n=1.4$ requires $\beta\approx 0.35$, which is consistent with the above results. A similar non-calorimetric model for the spatially resolved RC--SFR relation was proposed by \citet{murgia_05a}, who studied the molecular connection to the RC--FIR relation. \citet{paladino_06a} used it to
explain the observed spectral index behaviour in M\,51.

\subsection{Cosmic-ray transport}
\label{dis:cosmic_ray_transport}
The results of this study may provide several new insights to the diffusion process of cosmic rays using the CREs as proxy. In this section, we discuss limitations to our approach. Our inherent assumption is that the CRE timescale and thus the diffusion length is given by the CRE radiation lifetime as set by the synchrotron and IC losses. Clearly this is a simplification since galaxies are not electron calorimeters and the escape of CREs in winds has to be taken into account. With regards to the three galaxies we explored, we find that NGC~4736 has very low diffusion lengths and coefficients. Since the spectral indices in this galaxy are comparable to the injection index, spectral ageing seems to be suppressed because the CREs can quickly escape in a wind. We can presume that the escape time has to be smaller than the radiation lifetime of 24~Myr at 1365~MHz. Assuming a typical halo magnetic field scale height of 3~kpc \citep{heesen_18a}, we can calculate the minimum wind speed as $\varv=(3~{\rm kpc})/(24~{\rm Myr})=120~\rm km\,s^{-1}$. This is less than the escape velocity in this galaxy ($\varv_{\rm esc}=220~\rm km\,s^{-1}$) and therefore a plausible value for a minimum speed. We also notice that this galaxy has the highest \sfrd\  value in our sample and thus a fast wind is not surprising. Indeed, using the best-fitting correlation from \citet{heesen_18a}, we would expect a wind speed of $280~\rm km\,s^{-1}$, limiting the escape time further. Also, this galaxy has the smallest radius in our sample, and therefore the CREs may also escape across the outer boundary. A radial wind could support the action of the large-scale dynamo process and explain a coherent pattern of magnetic fields with unusually high radial component, which was observed in this object \citep{chyzy_08a}.

On the other hand, the other two galaxies, NGC~3184 and 5194, show signs of spectral ageing across their discs, in particular NGC~5194. In this galaxy, we find no energy dependence of the diffusion coefficient, whereas in NGC~3184 there is a mild dependence with $\mu=0.5$. In both galaxies we find diffusion coefficients that are slightly higher (factor of two), if using the analytical estimates (Section~\ref{subsec:cosmic_ray_diffusion_length}) or a factor of two or three lower when using the numerical estimates (Section~\ref{subsec:cosmic_ray_diffusion_model}), than the canonical value of the Milky Way of $3\times 10^{28}~\rm cm^2\,s^{-1}$ \citep{strong_07a}. These are encouraging results, although we would like to add more galaxies to improve our statistics. It is not unreasonable to assume that the diffusion coefficients are smaller than the Milky Way value, which is an average of the disc and halo coefficients. In the halo, the diffusion coefficient may be closer to $10^{29}~\rm cm^2\,s^{-1}$ \citep{liu_18a} and consequently in the disc the coefficient has to be smaller than the canonical value.

Similarly, the influence of the halo might explain the tension with the results of \citet{mulcahy_16a}, who find an isotropic diffusion coefficient of $6.6\times 10^{28}~\rm cm^2\,s^{-1}$ in NGC~5194, whereas we find a smaller value by a factor of four. We notice that this means only a difference of a factor of two in the diffusion lengths however. As \citet{mulcahy_16a} explained, a vertical escape was important to arrive at the observed spectral indices because otherwise the calculated spectral indices were too steep. So their spectral index was a fit to both the distribution in the disc as well as the vertical escape. It is conceivable that a vertical wind could also be present in this galaxy. In this case the diffusion coefficient may be smaller and explain some of the spectral index variations in the disc that are now smeared out by the large diffusion coefficients \citep[fig.~13 in][]{mulcahy_16a}.

\section{Conclusions}
\label{sec:conclusions}
We observed four nearby galaxies from the WSRT--SINGS survey \citep{braun_07a} with LOFAR, using 8 h on-source integrations to create deep 140 MHz maps. These maps were converted to spatially resolved distributions of the radio \sfrd, the SFR surface densities, using the widely-used relation by \citet{condon_92a}. These maps were compared with state-of-the art star formation tracers, the hybrid maps of \emph{GALEX} 156 nm FUV, and \emph{Spitzer} 24 $\mu$m MIR emission of \citet{leroy_08a}. Thus, we are able to calibrate the RC--SFR relation in nearby galaxies, removing any AGN contamination that can influence spatially unresolved studies. Planned wide-field surveys with the SKA and precursors/pathfinders in the 1--2~GHz range aim to probe the evolution of star formation across cosmic times. Since 140~MHz is the red-shifted rest-frame frequency of a galaxy emitting at $1.4$~GHz at redshift $z\approx 10$, it is useful to check whether the RC--SFR relation holds down to 140~MHz at redshift $z=0$ -- something we can test in nearby galaxies. These are our main results:

   \begin{enumerate}
      \item The radio \sfrd\  maps can be described as the smeared out versions of the hybrid \sfrd\  maps. This means that in the spiral arms or in other areas of active star formation the ratio of radio to hybrid \sfrd\ has a local minimum, whereas in the inter-arm regions and galaxy outskirts the ratio increases. For the LOFAR 140 MHz maps this effect is even more pronounced than for the WSRT 1365 MHz maps. This is expected, since the LOFAR maps show CREs that are a factor of three older than the WSRT maps if spectral ageing is important.
      \item A spatially resolved comparison of radio and hybrid \sfrd\  values averaged in regions of $1.2\times 1.2~\rm kpc^2$ show that the spatially resolved RC--SFR relation is sublinear with a slope of $a=0.59\pm 0.13$ for LOFAR and $a=0.75\pm 0.10$ for WSRT, where $(\Sigma_{\rm SFR})_{\rm RC}\propto [(\Sigma_{\rm SFR})_{\rm hyb}]^a$. This is another consequence of CRE diffusion, smearing out the RC emission, leading to this behaviour of the radio \sfrd\   maps.
      \item Within an individual galaxy, areas with young CREs with a radio spectral index of $\alpha>-0.65$, which is similar to the CRE injection spectral index, are in agreement with Condon's relation, whereas areas with steeper spectral indices tend to be radio bright. When plotting together our four sample galaxies, we found that areas with young CREs show a spatially resolved RC--SFR relation that is almost linear with $a\approx 0.9$, which is a unique relation independent of the hybrid \sfrd\  value.
      \item We convolved the hybrid \sfrd\  maps with a Gaussian kernel to simulate cosmic-ray diffusion and thus linearised the spatially resolved RC--SFR relation to measure diffusion lengths. We found that in two of the three galaxies, we tested this for, NGC~3184 and 5194, the cosmic-ray diffusion coefficient is $(0.9$--$1.5)\times 10^{28}~\rm cm^2\,s^{-1}$ at 1~GeV with no or a mild dependence on CRE energy ($\propto E^{0.5}$). The other galaxy, NGC~4736, clearly shows no spectral ageing and therefore our assumption of a steady state between injection and radiation losses may not be fulfilled. Hence, the derived diffusion coefficient of $0.13\times 10^{28}~\rm cm^2\,s^{-1}$ at 1~GeV may be too small. In this galaxy, the CREs probably escape before they have time to diffuse far in the disc. The minimum required advection speed would be $120~\rm km\,s^{-1}$.
      \item The spatially resolved RC--SFR relation in areas with young CREs suggests that in late-type spiral galaxies the non-thermal RC--SFR relation has to work for the non-calorimetric case. The CREs in the spiral arms diffuse away but in a stationary state, equating CRE injection and energy losses, which allows for the establishment of energy equipartition. Spiral galaxies have radio and optical radii that are larger than the CRE diffusion lengths of 1--5~kpc, which are the stationary values between 140 and 1365~MHz. Clearly, this relation is non-thermal since for LOFAR as the thermal contribution is negligible at 1~kpc resolution (much larger than the size of an H\,{\sc ii} region). A $B$--$\Sigma_{\rm SFR}$ and thus a $B$--$\rho^{\beta}$ relation is then the second requirement of such a non-calorimetric relation. Our results suggest $\beta\approx 0.4$, which is in approximate agreement with turbulent amplifications of magnetic fields.
      \item Our work emphasises the potential of using $140$-MHz RC observations as a star formation tracer out to a redshift of $z \approx 10$, in an equivalent way as we use $1.4$ GHz data in the local Universe.
   \end{enumerate}
   
   
Our results are very promising and show the potential of LOFAR observations of nearby galaxies. Clearly, the size of our sample is too small to draw statistically robust conclusions or search for trends with fundamental galaxies. In the near future, we expect to use further observations from the LoTSS to build up a larger sample, comparable to the one of \citet{heesen_14a} who studied 17 galaxies from WSRT--SINGS. These observations are now well underway and will also include edge-on galaxies allowing us to study the cosmic-ray transport in the haloes of spiral galaxies, comparing our maps with GHz observations \citep{irwin_12a}.

\begin{acknowledgements}
We thank the anonymous referee for an insightful and helpful report. LOFAR, the LOw Frequency ARray designed
and constructed by ASTRON, has facilities in several countries, which are owned
by various parties (each with their own funding sources) and are
collectively operated by the International LOFAR Telescope (ILT) foundation
under a joint scientific policy. This research was performed in the framework of the DFG Forschergruppe 1254 Magnetisation of Interstellar and Intergalactic Media: The Prospects of Low-Frequency Radio Observations. The data used in this work were in part processed on the Dutch national e-infrastructure with the support of SURF Cooperative through grant e-infra 160022 \& 160152. BNW acknowledges support from the Polish National Centre of Sciences (NCN), grant no. UMO-2016/23/D/ST9/00386. PNB is grateful for support from the UK STFC via grant ST/M001229/1. EB, CJ, LP, JW, and ES were supported by the National Science Foundation under grant  OISE 1458445, `Measuring Cosmic Magnetism with the Low Frequency Radio Array'.
\end{acknowledgements}


\bibliographystyle{aa}
\bibliography{sfr_low} 

\appendix


\section{Galaxy maps}
\label{app:maps}
\begin{figure*}
\centering
\includegraphics[width=\hsize]{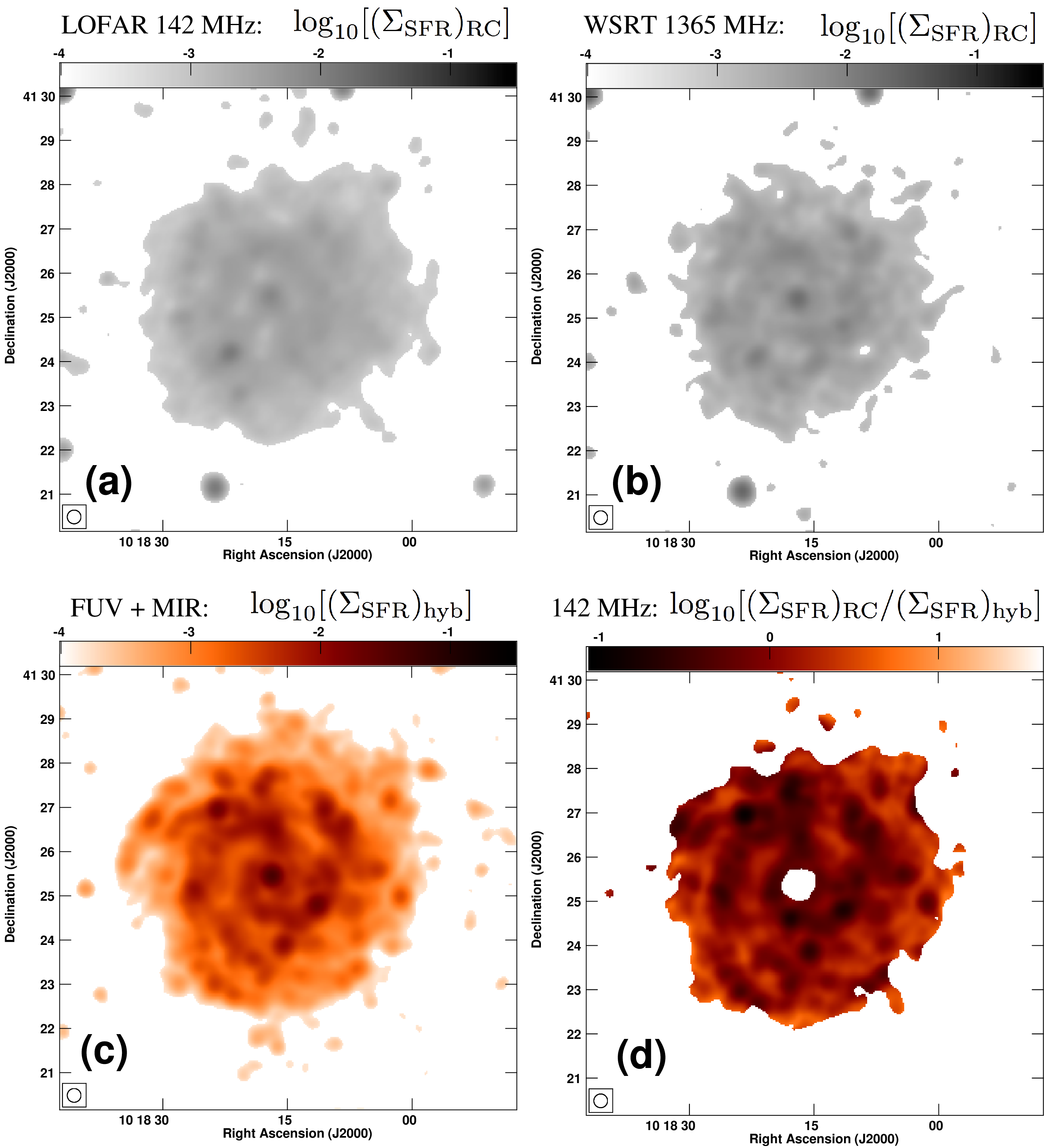} 
\caption{NGC~3184. (a) RC emission at 142~MHz, as derived from LOFAR HBA observations. The intensities were converted into a map of the radio $\Sigma_{\rm SFR}$ map, using the $1.4$ GHz relation of Condon scaled with a radio spectral index of $-0.8$. This map is shown at a logarithmic stretch ranging from $10^{-4}$ to $3\times 10^{-1}~\rm M_{\sun}\,yr^{-1}\,kpc^{-2}$. (b) Same as (a), but using a $1365$ MHz map from WSRT--SINGS. (c) Hybrid $\Sigma_{\rm SFR}$ map, derived from a linear superposition of \emph{GALEX} 156 nm FUV and \emph{Spitzer} 24 $\mu$m MIR emission, presented as inverted heat colour scale. (d) Ratio of the LOFAR $(\Sigma_{\rm SFR})_{\rm RC}$ map divided by the hybrid $(\Sigma_{\rm SFR})_{\rm hyb}$ map. The map is shown at logarithmic stretch using the heat colour scale ranging from $10^{-0.6}$ to $10^{1.6}$. Areas that are light are radio bright, whereas dark areas are radio dim when compared with the hybrid $\Sigma_{\rm SFR}$ map. All maps have been convolved to a circular Gaussian beam with a resolution of $18.6\times 18.6$~arcsec$^2$. The representation of the beam is shown in the bottom left corner of each panel. Panels (a)--(c) show unmasked maps, whereas panel (d) shows the area after masking background sources and the AGN-contaminated central area. In all panels, a 3$\sigma$ cut-off has been applied.}
\label{fig:n3184_lofar}
\end{figure*}

\begin{figure*}
\centering
\includegraphics[width=\hsize]{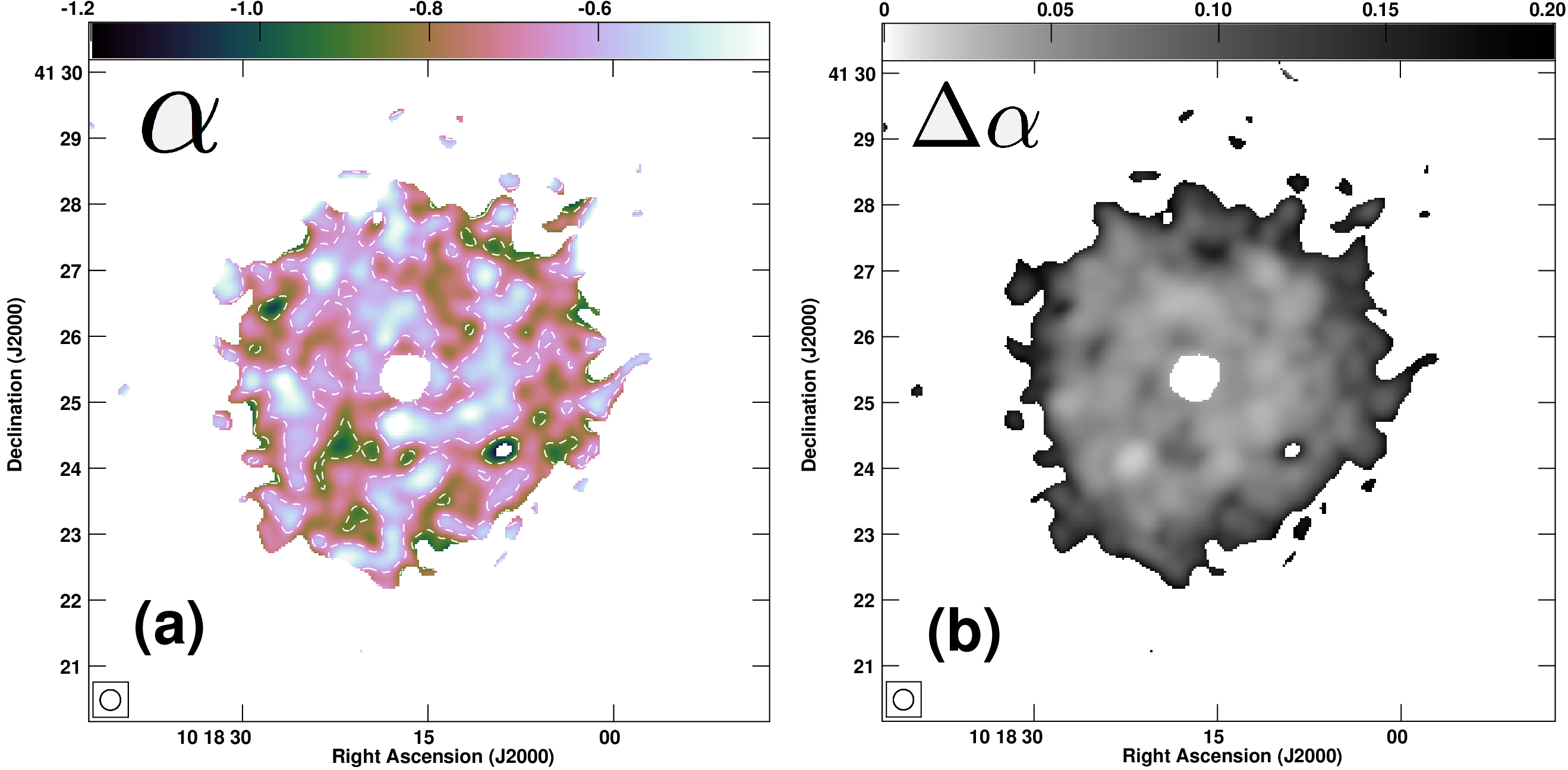} 
\caption{NGC~3184. (a) Radio spectral index distribution between 142 and 1365~MHz, presented in a cube-helix colour-scale ranging from $-1.2$ to $-0.4$. Dashed contours are at $-0.85$ and $-0.65$, thus separating the galaxy in three, non-coherent areas. Areas with young CREs ($\alpha > -0.65$) are predominantly found in spiral arms; areas with CREs of intermediate age ($-0.85 \leq \alpha \leq -0.65$) are predominantly found in inter-arm regions; and areas with old CREs ($\alpha < -0.85$) are found in the galaxy's outskirts. (b) Error of the radio spectral index distribution between 145 and 1365~MHz, presented in a cube-helix colour-scale ranging from 0 to $0.2$. As can bee seen, only in areas of steep spectral indices $\alpha < -0.85$ the spectral index error becomes larger than $\pm 0.1$. In both panels, the maps were convolved to a circular synthesised beam of $18.6\times 18.6~\rm arcsec^2$ resolution, which is outlined in the bottom left corner. A mask has been applied to background sources and the central regions of NGC~3184. A 3$\sigma$ cut-off was applied to both the 142- and 1365-MHz maps prior to combination.}
\label{fig:n3184_spix}
\end{figure*}

\begin{figure*}
\centering
\includegraphics[width=\hsize]{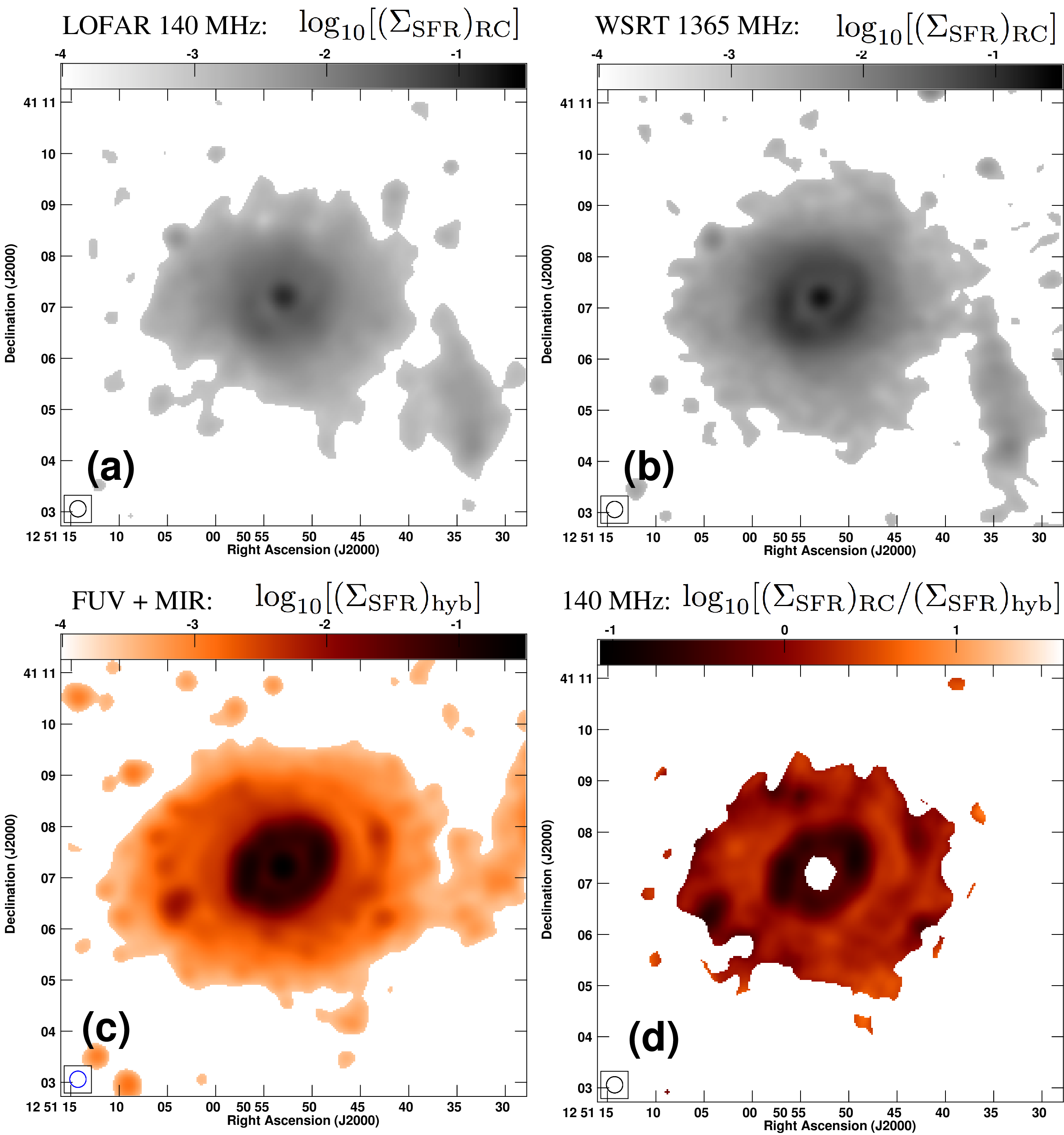} 
\caption{NGC~4736. (a) RC emission at 140~MHz, as derived from the LOFAR HBA observations. The intensities were converted into a map of the radio $\Sigma_{\rm SFR}$ map, using the $1.4$ GHz relation of Condon scaled with a radio spectral index of $-0.8$. This map is shown at a logarithmic stretch ranging from $10^{-4}$ to $3\times 10^{-1}~\rm M_{\sun}\,yr^{-1}\,kpc^{-2}$. (b) Same as (a), but using a $1365$ MHz map from WSRT--SINGS. (c) Hybrid $\Sigma_{\rm SFR}$ map, derived from a linear superposition of \emph{GALEX} 156 nm FUV and \emph{Spitzer} 24 $\mu$m MIR emission, presented as inverted heat colour scale. (d) Ratio of the LOFAR $(\Sigma_{\rm SFR})_{\rm RC}$ map divided by the hybrid $(\Sigma_{\rm SFR})_{\rm hyb}$ map. The map is shown at logarithmic stretch using the heat colour scale, ranging from $10^{-0.6}$ to $10^{1.6}$. Areas that are light are radio bright, whereas dark areas are radio dim when compared with the hybrid $\Sigma_{\rm SFR}$ map. All maps have been convolved to a circular Gaussian beam with a resolution of $19.1\times 19.1$~arcsec$^2$. The representation of the beam is shown in the bottom left corner of each panel. Panels (a)--(c) show unmasked maps, whereas panel (d) shows the area after masking background sources and the AGN-contaminated central area. In all panels, a 3$\sigma$ cut-off has been applied.}
\label{fig:n4736_lofar}
\end{figure*}

\begin{figure*}
\centering
\includegraphics[width=\hsize]{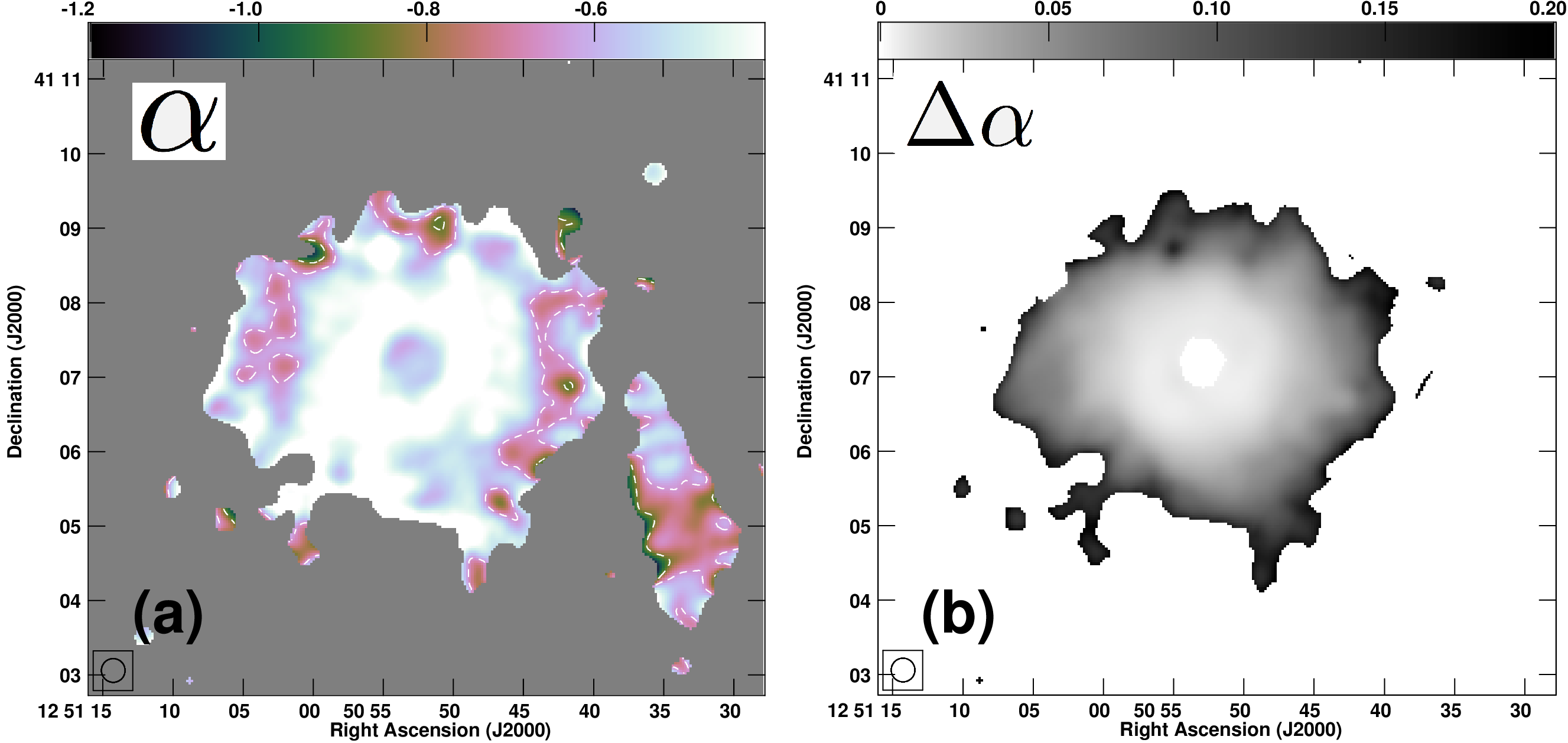} 
\caption{NGC~4736. (a) Radio spectral index distribution between 140 and 1365~MHz, presented in a cube-helix colour scale ranging from $-1.2$ to $-0.4$. Dashed contours are at $-0.85$ and $-0.65$, thus separating the galaxy into three non-coherent areas. Areas with young CREs ($\alpha > -0.65$) are predominantly found in spiral arms; areas with CREs of intermediate age ($-0.85 \leq \alpha \leq -0.65$) are predominantly found in inter-arm regions; and areas with old CREs ($\alpha < -0.85$) are found in the galaxy outskirts. (b) Error of the radio spectral index distribution between 145 and 1365~MHz, presented in a cube-helix colour scale ranging from 0 to $0.2$. As can be seen, the spectral index error only becomes larger than $\pm 0.1$  in
areas of steep spectral indices $\alpha < -0.85$. In both panels, the maps were convolved to a circular synthesised beam of $19.1\times 19.1~\rm arcsec^2$ resolution, which is outlined in the bottom left corner. A mask has been applied to background sources and the central regions of NGC~4736. A 3$\sigma$ cut-off was applied to both the 142 and 1365 MHz maps prior to combination.}
\label{fig:n4736_spix}
\end{figure*}

\begin{figure*}
\centering
\includegraphics[width=\hsize]{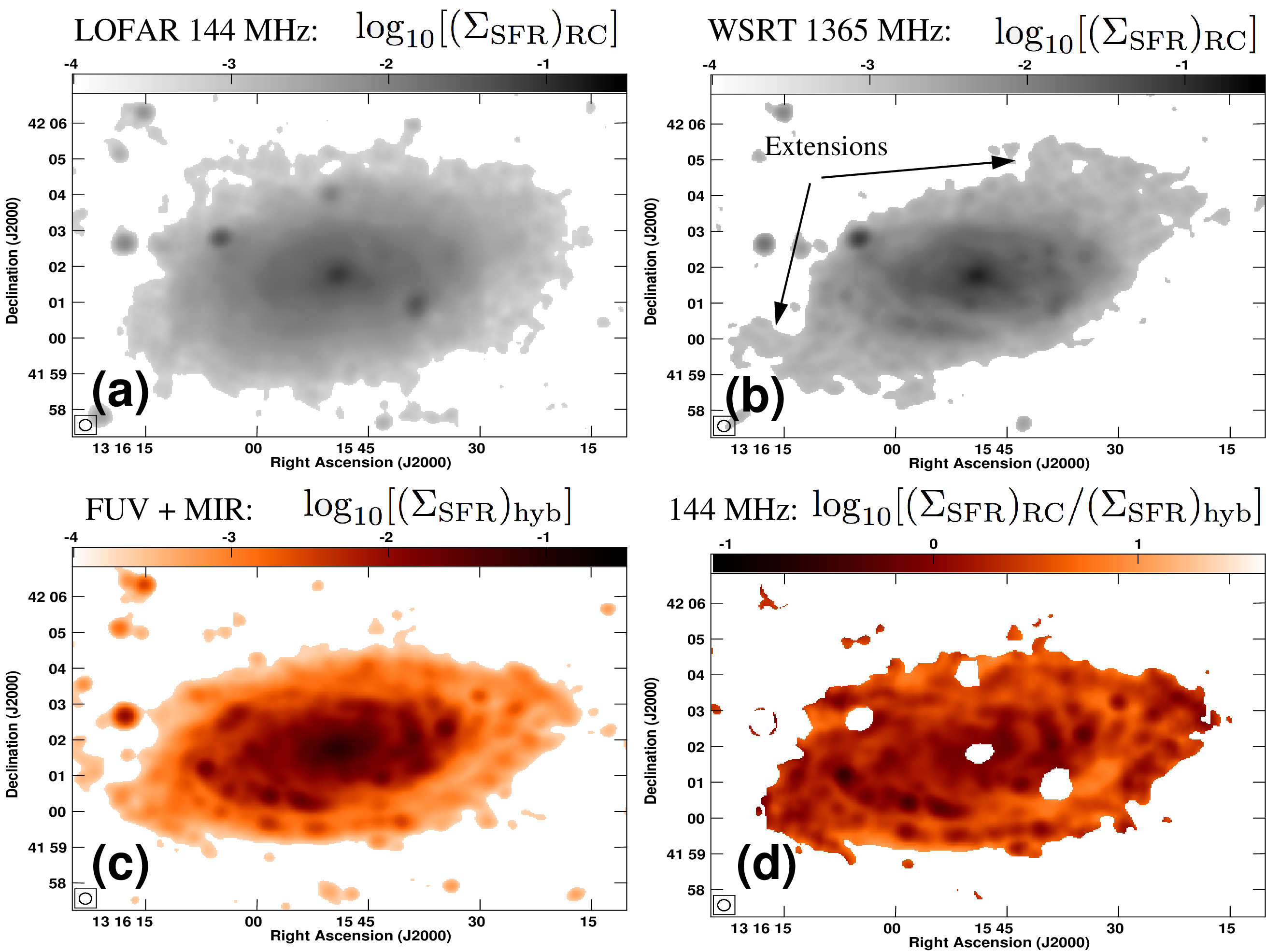} 
\caption{NGC~5055. (a) RC emission at 144~MHz, as derived from the LOFAR HBA observations. The intensities were converted into a map of the radio $\Sigma_{\rm SFR}$ map, using the $1.4$ GHz relation of Condon scaled with a radio spectral index of $-0.8$. This map is shown at a logarithmic stretch ranging from $10^{-4}$ to $3\times 10^{-1}~\rm M_{\sun}\,yr^{-1}\,kpc^{-2}$. (b) Same as (a), but using a $1365$ MHz map from WSRT--SINGS. (c) Hybrid $\Sigma_{\rm SFR}$ map, derived from a linear superposition of \emph{GALEX} 156 nm FUV and \emph{Spitzer} 24 $\mu$m MIR emission, presented as inverted heat colour scale. (d) Ratio of the LOFAR $(\Sigma_{\rm SFR})_{\rm RC}$ map divided by the hybrid $(\Sigma_{\rm SFR})_{\rm hyb}$ map. The map is shown at logarithmic stretch using the heat colour scale ranging from $10^{-0.6}$ to $10^{1.6}$. Areas that are light are radio bright, whereas dark areas are radio dim when compared with the hybrid $\Sigma_{\rm SFR}$ map. All maps have been convolved to a circular Gaussian beam with a resolution of $19.1\times 19.1$~arcsec$^2$. The representation of the beam is shown in the bottom left corner of each panel. Panels (a)--(c) show unmasked maps, whereas panel (d) shows the area after masking background sources and the AGN-contaminated central area. In all panels, a 3$\sigma$ cut-off has been applied.}
\label{fig:n5055_lofar}
\end{figure*}

\begin{figure*}
\centering
\includegraphics[width=\hsize]{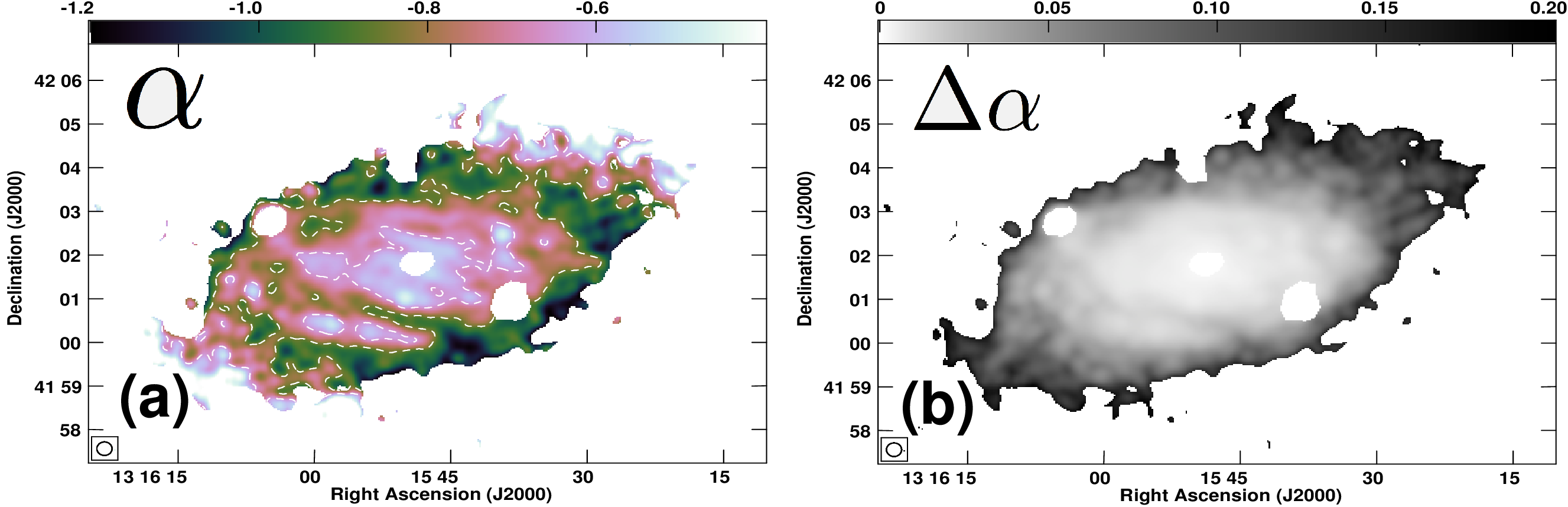} 
\caption{NGC~5055. (a) Radio spectral index distribution between 140 and 1365~MHz, presented in a cube-helix colour scale ranging from $-1.2$ to $-0.4$. Dashed contours are at $-0.85$ and $-0.65$, thus separating the galaxy into three non-coherent areas. Areas with young CREs ($\alpha > -0.65$) are predominantly found in spiral arms; areas with CREs of intermediate age ($-0.85 \leq \alpha \leq -0.65$) are predominantly found in inter-arm regions; and areas with old CREs ($\alpha < -0.85$) are found in the outskirts of the galaxy. (b) Error of the radio spectral index distribution between 145 and 1365~MHz, presented in a cube-helix colour scale ranging from 0 to $0.2$. As can be seen, the spectral index error only becomes larger than $\pm 0.1$ in areas of steep spectral indices $\alpha < -0.85$. In both panels, the maps were convolved to a circular synthesised beam of $19.1\times 19.1~\rm arcsec^2$ resolution, which is outlined in the bottom left corner. A mask has been applied to background sources and the central regions of NGC~5055. A 3$\sigma$ cut-off was applied to both the 142 and 1365 MHz maps prior to combination.}
\label{fig:n5055_spix}
\end{figure*}

In this on-line-only appendix, we present the remaining galaxies NGC~3184, 4736, and 5055, the maps of which were not shown in the main paper. Figs~\ref{fig:n3184_lofar}, \ref{fig:n4736_lofar}, and \ref{fig:n5055_lofar} show the radio \sfrd\  maps from LOFAR and WSRT observations as well as the hybrid \sfrd\  maps from combined \emph{GALEX} 156 nm FUV and \emph{Spitzer} 24~$\mu$m MIR observations. We also present the ratio of the LOFAR \sfrd\  map to the hybrid \sfrd\  map. These maps are shown at logarithmic stretch. In Figs~\ref{fig:n3184_spix}, \ref{fig:n4736_spix}, and \ref{fig:n5055_spix}, we present the spectral index maps between LOFAR 140~MHz and WSRT 1365~MHz together with a map of the spectral index uncertainties.

The WSRT maps are available on the NASA Extragalactic Database (NED) and are from \citet{braun_07a}. The LOFAR RC maps will be made available by us on the website of the Centre de Donn\'ees astronomiques de Strasbourg (CDS) (\href{http://cds.u-strasbg.fr}{http://cds.u-strasbg.fr}).

\end{document}